\documentclass[sigconf]{acmart}

\usepackage[english]{babel}
\usepackage{blindtext}

\renewcommand\footnotetextcopyrightpermission[1]{}
\setcopyright{none}

\hypersetup{
    colorlinks=true,
    linkcolor=blue,
    filecolor=magenta,
    urlcolor=cyan,
}

\acmDOI{}

\acmISBN{}

\acmPrice{}

\usepackage{amsmath}
\ifcsname Bbbk\endcsname
  
\fi
\usepackage{amssymb}
\usepackage{booktabs}

\usepackage{lineno}
\usepackage{pifont}
\usepackage{algorithm}
\usepackage[noend]{algpseudocode}

\usepackage{algpseudocode}
\usepackage{float}
\usepackage{subcaption}
\usepackage{graphicx}
\usepackage{textcomp}
\usepackage{xcolor}
\usepackage{tabularx}
\usepackage{xspace}
\usepackage{breakurl}
\usepackage{url}
\usepackage{multirow}
\usepackage{pifont}
\usepackage[inline]{enumitem}
\usepackage{tikz}
\usepackage{comment}
\usepackage{amsthm}

\usepackage{makecell}
\usepackage[normalem]{ulem}
\usepackage{hyperref}

\renewcommand\footnotemark{}

\newcommand{\secref}[1]{\hyperref[#1]{\S\ref*{#1}}}
\newcommand{\figref}[1]{\hyperref[#1]{Figure~\ref*{#1}}}
\newcommand{\tabref}[1]{\hyperref[#1]{Table~\ref*{#1}}}
\newcommand{\algoref}[1]{\hyperref[#1]{Algorithm~\ref*{#1}}}

\newcommand{\paraspace}{\vspace{0.03in}}
\newcommand{\parab}[1]{\paraspace\noindent{\bf #1} }

\newcommand{\sys}{{Nuska}\xspace}
\newcommand{\DAP}{\emph{Defer-and-Promote}\xspace}

\newcommand{\eg}{e.g., }

\algnewcommand{\LineComment}[1]{\Statex \hspace{\algorithmicindent}\textcolor{gray}{\(\triangleright\) #1}}
\algnewcommand{\FuncInput}[1]{\Statex \textcolor{black}{\(\triangleright\) \textbf{input}\xspace#1}}
\algnewcommand{\FuncOutput}[1]{\Statex \textcolor{black}{\(\triangleright\) \textbf{output}\xspace#1}}

\let\oldding\ding
\renewcommand{\ding}[2][1]{\scalebox{#1}{\oldding{#2}}}

\begin{document}

\pagestyle{plain}

\title{Nuska: Stage-Aware Communication Scheduling for Disaggregated LLM Serving}

\author{
  Yijun Sun$^{1}$ \quad
  Xudong Liao$^{1}$ \quad
  Songrun Xie$^{1}$ \quad
  Hao Chen$^{2}$ \quad
  Han Tian$^{3}$ \\
  Wenxue Li$^{1}$ \quad
  Yiming Zhang$^{2}$ \quad
  Kai Chen$^{1}$ \\
  \vspace{0.2cm}
  $^{1}$iSING Lab, Hong Kong University of Science and Technology \\
  $^{2}$Shanghai Jiao Tong University \quad
  $^{3}$University of Science and Technology of China \\
  \vspace{0.1cm}
}

\renewcommand{\shortauthors}{Sun et al.}

\begin{abstract}
Meeting stringent Time-To-First-Token (TTFT) requirements is crucial for LLM applications. To improve efficiency, modern LLM serving systems adopt disaggregated architectures with diverse parallelisms, introducing complex multi-stage workflows involving reusable KV-block retrieval, collective communication, and P2D transfer. Flows from dependent stages overlap within and across requests on shared bottleneck links, making TTFT highly susceptible to network contention and necessitating stage-aware scheduling. Unfortunately, most existing works schedule flows in a stage-agnostic manner, leading to uncoordinated contention that constitutes a primary cause of SLO violations.

In this paper, we present \sys, a holistic multi-stage flow scheduling mechanism designed to maximize TTFT SLO attainment. At its core, \sys approximates the Least-Laxity-First (LLF) scheduling policy without requiring precise knowledge of a request's remaining slack. It achieves this through a \emph{Defer-and-Promote} principle implemented through a Reverse Multi-Level Queue (RMLQ) structure. By dynamically promoting task precedence as effective laxity diminishes, \sys prioritizes flows with less laxity while preventing requests with loose SLOs from prematurely consuming network bandwidth. We implement \sys as a pluggable module integrated into vLLM, and evaluate it on a 8-server, 32-GPU testbed as well as through large-scale simulations. Our results demonstrate that \sys effectively outperforms state-of-the-art baselines, improving the TTFT SLO attainment by 1.2$\times$--2.4$\times$.

\end{abstract}

\maketitle

\section{Introduction}
The rapid advancement of large language models (LLMs) has spawned a wide range of use cases, including interactive chatbots~\cite{zheng2023judging,chiang2023vicuna,openai2024gpt4}, code assistants~\cite{liu2024exploring,zhong2024can}, and agentic systems~\cite{talebirad2023multi,he2025llm,yu2024fincon}.
Modern serving systems must sustain massive request volumes while adhering to heterogeneous service-level objectives (SLOs)~\cite{tempo2025,patke2024queue,hongsola,chen2025slos,shen2025accelgen} to meet stringent performance expectations.
Among these metrics, time-to-first-token (TTFT) remains particularly critical, as it governs both the responsiveness required for interactive users~\cite{wang2024revisiting,ethiraj2025toward} and the strict timing constraints of agentic pipelines~\cite{LiteLLM,langfuse,notdiamond}.
Violations of TTFT targets often trigger repetitive request retries~\cite{Medium1,Medium2}, leading to unnecessary resource consumption~\cite{mooncake} and significant revenue loss~\cite{ranganathan2025enhancing}.

\begin{figure}[t!]
    \centering
    \includegraphics[width=0.9\linewidth]{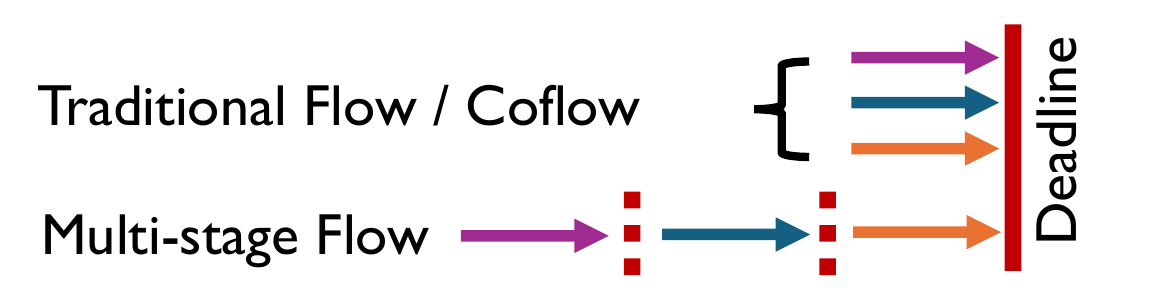}
    \caption{Illustration of difference between Multi-stage Flow Scheduling and prior stage-agnostic scheduling.}
    \vspace{-1.5em}
    \label{fig:intro:MFS}
\end{figure}

Modern LLM serving systems typically adopt disaggregated prefill–decode architectures with KV-cache reuse~\cite{distserve,splitwise,strati2024d,hu2024inference,mooncakefast,hu2024memserve,cachedattention,cacheblend} and leverage diverse forms of  parallelism~\cite{expertparallel,deepspeedMoE,deepseektech,comet} to improve efficiency. Despite their promise, they still suffer from substantial communication overhead during the prefill phase. Specifically, generating the first token involves \emph{multi-stage} communication: (1) remote retrieval of reusable KV-blocks, (2) collective communication for the synchronization of intermediate activations, and (3) Prefill-to-Decode (P2D) transfer. Flows from different stages frequently overlap and contend for shared network fabrics, which leads to \emph{intra-request contention}, where different communication phases within the same request interfere with each other, and \emph{inter-request contention}, where communication from different batches or prefill units competes for shared links. Such contention significantly inflates prefill TTFT latency and heightens the risk of system-wide SLO violations.

Unfortunately, prior works fail to meet end-to-end deadlines under contention (Figure~\ref{fig:intro:MFS}), largely because they treat flow in stage-agnostic manner (i.e., without awareness of their stage dependencies). Traditional flow scheduling schemes (e.g., Fair Sharing~\cite{dctcp,dcqcn}, SJF~\cite{pfabric,PDQ,pias,li2024flow}, EDF~\cite{D3,d2tcp,PDQ}, etc.) operate on individual flows without holistic view. Even coflow-based approaches~\cite{varys,Aalo,zhang2016coda,zhao2015rapier,agarwal2018sincronia,wan2025coflow} fall short: while they manage data-parallel flows as a group, they ignore the inherent dependencies across stages. Consequently, these schemes may over-prioritize early-stage communication or starve latency-critical downstream stages, ultimately degrading TTFT SLO attainment.

To this end, we ask: \emph{Can we schedule multi-stage communication holistically in LLM serving systems to maximize overall TTFT SLO attainment?}

When exploring the design space, we note distinct efforts that optimize isolated stages, targeting either collective communication~\cite{rajasekaran2024cassini,cao2024crux,xu2025autoccl,cao2025syccl,si2025collective} or KV-cache transfers~\cite{strati2024d,cachegen,chen2024kvdirect,mooncakefast}. However, these specialized solutions remain insufficient. The reason is that they often pursue conflicting optimization objectives, naively combining them can inadvertently amplify network contention.

We answer this question affirmatively with \sys{}, a \emph{holistic} multi-stage communication scheduler that jointly orchestrates dependent communication stages to maximize TTFT SLO attainment. Our key insight is that the global TTFT deadline can be gradually translated into explicit flow-level deadlines as prefill execution progresses. Leveraging this insight, \sys{} approximates the Least-Laxity-First (LLF) policy under uncertainty and realizes a \emph{Defer-and-Promote} principle via a Reverse Multi-Level Queue (RMLQ). By dynamically promoting task precedence based on diminishing effective slack, \sys{} prioritizes genuinely critical stages while preventing requests with loose SLOs from prematurely consuming network bandwidth.

Although our insight is straightforward,
translating it into an efficient, real-world system requires addressing several non-trivial challenges. First, how to schedule last-stage communications with explicit deadline without incurring excessive prioritization? Second, how to schedule early-stage flows governed by implicit TTFT bound under uncertain laxity? Third, how to ensure \sys{}’s compatibility with emerging collective communication and KV-cache libraries?

For the first challenge, we employ a lazy promotion strategy leveraging Minimal Link Utilization (MLU) as the trigger. P2D flows initialize in low-priority queues and are promoted only when their MLU exceeds a predefined threshold. This ensures that P2D flows receive only the minimum bandwidth for just-in-time completion, preserving bandwidth headroom for early-stage communications. To avoid priority thrashing, promotions occur only at layer granularity. This approach enables effective coarse-grained  prioritization while facilitating  practical implementation on commodity switches with hardware priorities queues and lightweight packet tagging, thereby eliminating the risk of packet re-ordering.

To address the second issue, we utilize a two-tier scheduling strategy. For intra-request contention within a single prefill unit, we prioritize collective communication that unblock subsequent computation, while elevating KV-cache transfers only when they threaten to stall the next layer's execution. For inter-request contention across the prefill cluster, we order units based on TTFT deadlines and apply feasibility checks to prune infeasible communication tasks to avoid blocking. This design prevents premature promotion when laxity is unclear, while keeping early-stage communication aligned with downstream TTFT requirements.

Finally, to ensure compatibility, \sys{} integrates with existing stacks (e.g., NCCL \cite{nccl}, Mooncake \cite{mooncakeTE}) via lightweight task adapters. These adapters intercept tasks for precise host-side prioritization in software queues, while mapping priorities to DSCP values for robust traffic isolation on switches. By employing hybrid priority enforcement, \sys{} ensures seamless integration and effective contention resolution using standard network primitives.

We build a \sys prototype and evaluate it on a 8-server testbed with 32 NVIDIA 3090 GPUs \cite{nvidia3090}, 16 Mellanox NICs~\cite{cx5}, all connected to a single Top-of-Rack (ToR) switch. We implement \sys{} as a pluggable module into NCCL and Mooncake, and integrate it with the vLLM inference engine. Using this prototype, we successfully demonstrated the benefits of \sys{} on the state-of-the-art LLM~\cite{mixtral7b}.

To evaluate the performance of \sys{} at scale, we further perform large-scale simulations using four representative real-world MoE models~\cite{mixtral22b,grok,dbrx,qwen-coder}.
Our results reveal that \sys{} significantly outperforms state-of-the-art scheduling schemes, improving the TTFT SLO attainment rate by 1.2$\times$-2.4$\times$ compared to baselines.
We also observe that \sys{} cutting down non-overlapped collective completion time by 50$\%$ and also improving the request earliness greatly.

\section{Background and Motivation}\label{sec:background}

\subsection{LLM Serving and TTFT}

\begin{figure}[t!]
    \centering

    \includegraphics[width=\linewidth]{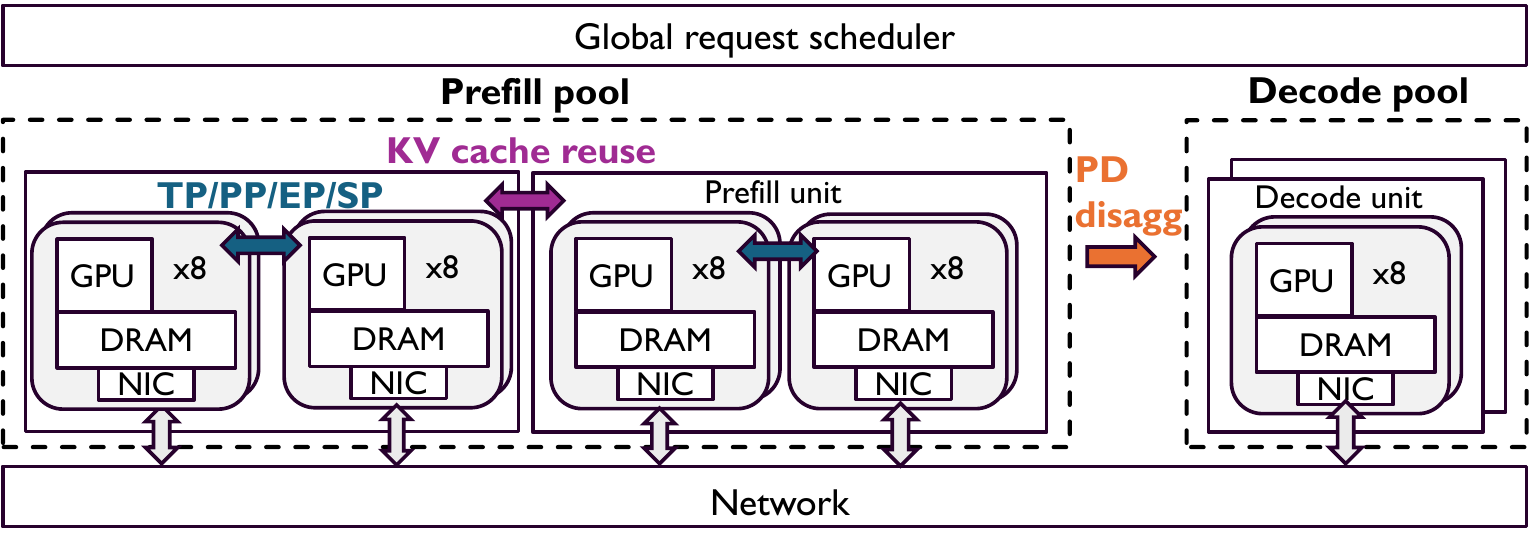}
    \vspace{-0.5em}
    \caption{Illustration of LLM serving at scale.}
    \label{fig:bg:servingoverview}

\end{figure}

Large language models have become the foundation of modern AI services. These models are inherently autoregressive: the system first processes the entire prompt to generate the first token (the prefill stage), and then produces subsequent tokens iteratively using the accumulated KV cache (the decode stage). Performance is typically characterized by Time-to-First-Token (TTFT) for the prefill phase and Time-Between-Tokens (TBT) for the decode phase.

TTFT is a critical service-level objective (SLO) for many LLM applications. In interactive scenarios like chatbots and voice assistants, users expect an immediate initial response, and some studies indicate noticeable disengagement when latency exceeds a few hundred milliseconds and abandonment beyond seconds~\cite{vijaya2025aqua,liu2025m,kim2025seconds}. TTFT is also vital for automatic agentic frameworks, which rely on strict timeouts~\cite{langfuse,LiteLLM} for fault tolerance. A delayed first token triggers costly remedial actions such as service retries~\cite{Medium1,Medium2} or provider switching~\cite{notdiamond}. A recent report from Microsoft~\cite{ranganathan2025enhancing} reveals that the delayed first token is one of the primary source ($\sim 40\%$) of service failure (reported as timeout error), leading to revenue loss, higher support costs, and reputational risk.

As depicted in Fig.~\ref{fig:bg:servingoverview}, modern production systems~\cite{mooncake,deepseektech} scale out to clusters comprising thousands of nodes interconnected with high-bandwidth links, where each node contains multiple xPUs (\eg GPUs, TPUs, NPUs). Nodes are organized into individual serving units, each dedicated to host one model replica. Models are deployed using a combination of tensor~\cite{narayanan2021efficient}, pipeline~\cite{narayanan2019pipedream, alphaserve}, sequence~\cite{ringattention,wu2024loongserve,CP}, and expert parallelism~\cite{expertparallel,deepspeedMoE,deepseektech,comet}. To further optimize efficiency, recent work~\cite{distserve,splitwise,strati2024d,hu2024inference}  proposes \emph{prefill–decode disaggregation}, which assigns prefill tasks to compute-optimized devices and decode tasks to memory-rich devices, thereby accommodating their heterogeneous resource demands. In addition, \emph{KV cache reuse}~\cite{mooncakefast,hu2024memserve,cachedattention,cacheblend} has been introduced to amortize prefill costs across requests by transferring cached contexts between nodes.

\begin{figure}[t!]
    \centering
    \includegraphics[width=\linewidth]{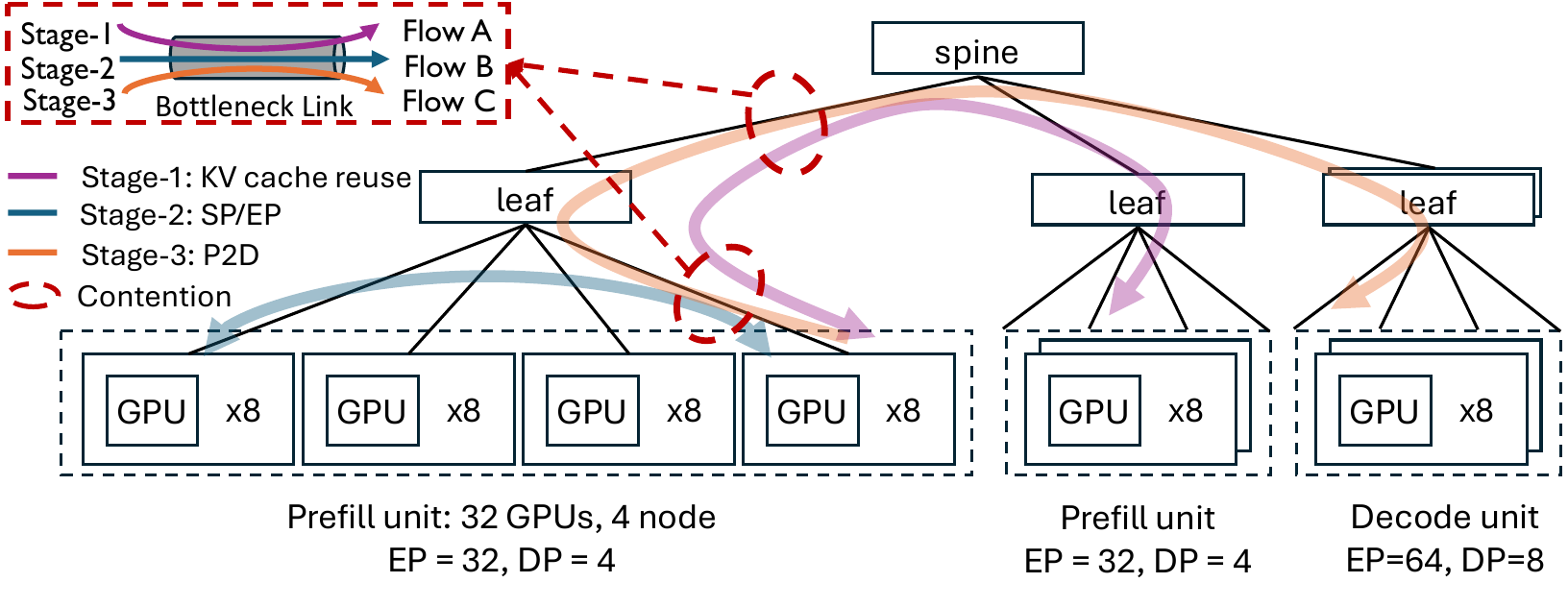}
    \vspace{-0.5em}
\caption{An example illustrating communication contention in LLM serving systems: KV cache transfers (Stage~1 \& 3) and parallelism communication (Stage~2) compete for the same link bandwidth.}
    \label{fig:bg:contentionmodel}
\end{figure}

\begin{figure*}[ht!]
    \centering

    \includegraphics[width=0.9\linewidth]{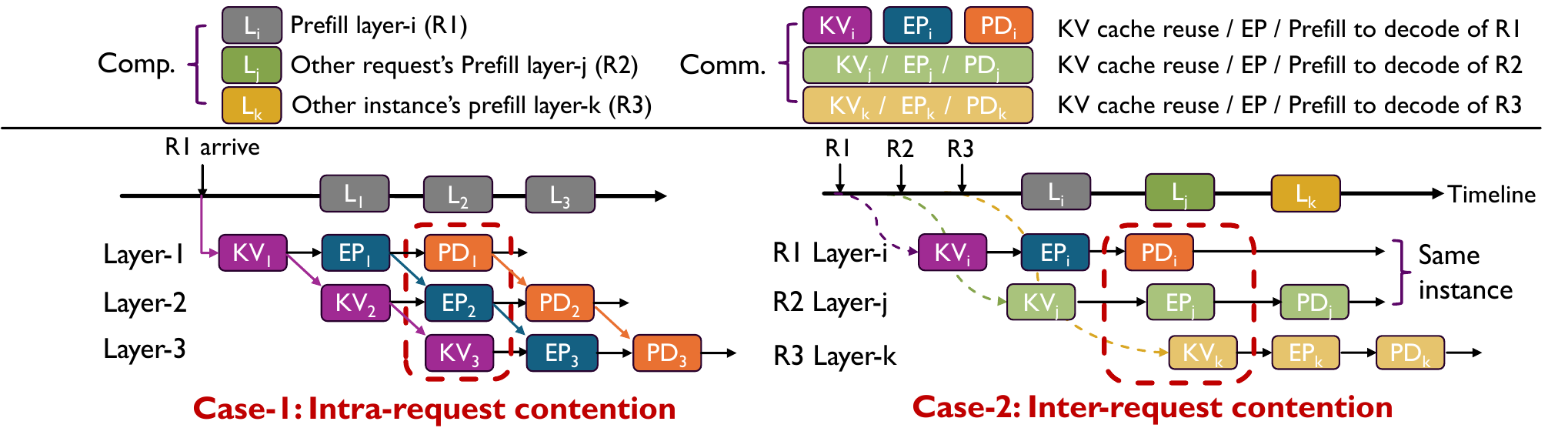}

    \caption{Illustration of two primary forms of contention in LLM serving systems.}
    \label{fig:bg:contention}
\end{figure*}

\subsection{Contention Across Multi-Stage Communication}

Generating the first token in existing production systems (Fig. \ref{fig:bg:servingoverview}) consists of multi-stage communications:
\begin{enumerate}[label=\(\bullet\),leftmargin=*]
    \item \textbf{Stage~1: KV-cache reuse}: fetches reusable KV-cache blocks from remote prefill unit;
    \item \textbf{Stage~2: Collective communication}: exchanges intermediate activations across devices via model parallelism;
    \item \textbf{Stage~3: Prefill-to-Decode (P2D) transfer}: delivers the full KV-cache history to the decode unit \footnote{Mainstream open-source serving frameworks~\cite{vllm_pd_2025,sglang_pd_2025,dynamo_pd_2025} explicitly incorporate Stage~3 latency into the TTFT metric.}.
\end{enumerate}

Prior literature has highlighted the substantial overhead of individual stages, noting that collective communication (e.g, all-to-all, etc.) occupy 40--60\% of end-to-end latency \cite{linaatc,comet,li2025optimizing,CP} and KV-cache movement contributes a non-trivial share~\cite{cachegen,mooncake,chen2024kvdirect,li2025flowkv,yoon2025tract}. In this paper, we further identify a critical yet underexplored problem: collective communication and KV-cache movement frequently overlap in time and contend for shared network bandwidth. Fig.~\ref{fig:bg:contentionmodel} illustrates the spatial location of contention: KV-cache transfers (Stage~1\&3) and collective communication (Stage~2) traverse shared physical interconnects. This spatial co-location creates two primary forms of contention, as shown in Fig.~\ref{fig:bg:contention}:
\begin{itemize}[leftmargin=*]
    \item \textbf{Intra-request contention} occurs when communication flows interfere within a single request. Specifically, layer-wise KV-cache transfers (Stage~1\&3) compete directly with the ongoing collective communication on other layers (Stage~2) for shared link bandwidth.
    \item \textbf{Inter-request contention} arises when concurrent requests from different prefill units interfere, resulting in frequent link competition in the switch links. Moreover, driven by skewed block popularity, multiple prefill units may converge on a single remote \textit{victim unit} to fetch ``hot'' KV blocks, causing contention between the victim unit's local collective communication (Stage~2) and remote KV-cache fetching on its NIC bandwidth (Stages~1\&3).
\end{itemize}

\begin{figure}[t]
    \centering
    \begin{subfigure}[t]{0.48\linewidth}
        \centering
        \includegraphics[width=\linewidth]{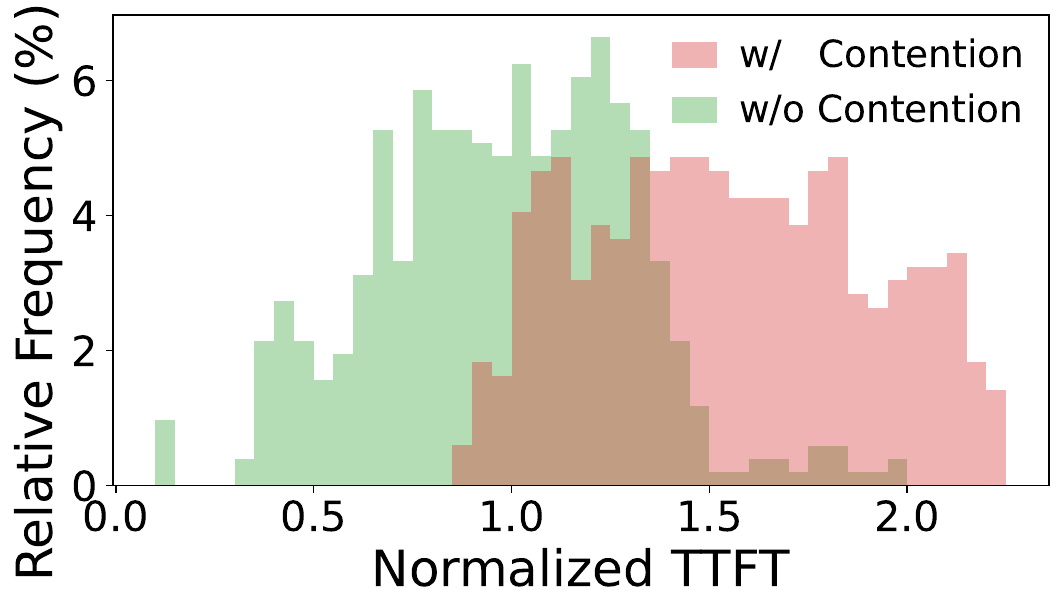}
        \caption{Impact of contention on end-to-end TTFT.}
        \label{fig:moti:ttft}
    \end{subfigure}
    \hfill
    \begin{subfigure}[t]{0.48\linewidth}
        \centering
        \includegraphics[width=\linewidth]{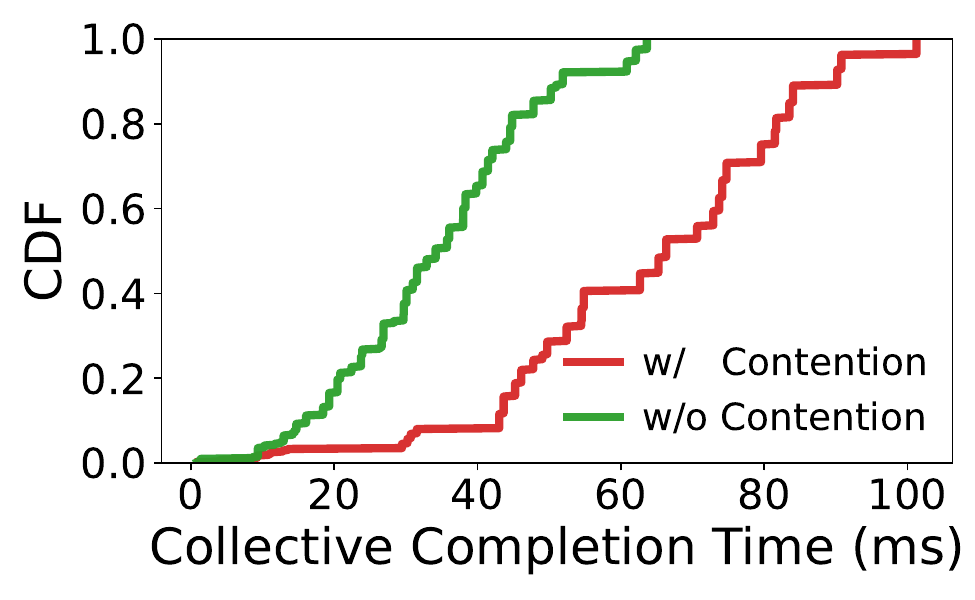}
        \caption{Impact of contention on all-to-all latency.}
        \label{fig:moti:cct}
    \end{subfigure}
    \hfill

    \caption{[Testbed] Impact of communication contention on Mixtral 8x7B model.}

    \label{fig:moti:contentionimpact}
\end{figure}

To quantify the performance degradation caused by such interference, we conduct measurements on a 16-GPU prefill cluster (50 Gbps/GPU) hosting two serving instances ($TP=1, EP=8$). We use QwenB-agent, an agent workload from production-derived Qwen traces~\cite{qwen_bailian_traces}, with average sequence length 1k tokens, 65\% prompt reuse, and per-GPU request rate 1 req/s; the full testbed setup is described in Sec.~\ref{sec:evaluation:testbed}. We evaluate the impact of contention on two metrics: end-to-end TTFT and collective communication time (CCT) of All-to-All operations (aggregating Dispatch and Combine phases). Specifically, we perform a comparative analysis between an ideal baseline (\textit{w/o contention}) and a realistic serving scenario (\textit{w/ contention}). Fig.~\ref{fig:moti:ttft} illustrates the TTFT distribution. The results reveal that, under contention, the overall TTFT is prolonged by nearly $50\%$. We attribute this inflation primarily to the latency degradation in All-to-All communication on the critical path, whose CCT nearly doubles ($1.8\times$) due to contention with concurrent KV-cache operations (as shown in Fig.\ref{fig:moti:cct}). This substantial slowdown confirms that network contention is a major source of performance variability in the prefill phase.

\subsection{Limitation of Existing Work}
The previous analysis identifies unmanaged multi-stage communication contention as a primary source of TTFT violations. We categorize existing approaches into two groups: recent system-level communication optimizations and conventional scheduling algorithms. As detailed below, they both fall short in coordinating such contention due to the lack of a holistic view of multi-stage dependencies.

\parab{Stage-agnostic flow scheduling.} Conventional datacenter scheduling disciplines manage flows or coflows individually to optimize network metrics, such as flow/coflow completion time or deadline satisfaction. However, these objectives are misaligned with end-to-end application goals (e.g., TTFT attainment), because TTFT is jointly determined by flows across multiple interdependent stages. Consequently, these approaches inevitably over-prioritize slack-rich flows while starving the critical transfers required to unlock downstream computation, ultimately degrading TTFT SLO attainment.

\parab{Specialized optimization for isolated stages.}
We notice some distinct efforts optimize isolated stages, specifically targeting collective communication via algorithm synthesis~\cite{shah2023taccl,liu2024rethinking,kim2024tccl,cao2025syccl}, parameter tuning~\cite{xu2025autoccl}, and overlapping~\cite{deepseektech,comet,zheng2025triton,zheng2025tilelink}, or KV-cache transfers via pipelined prefetching \cite{splitwise,mooncakefast} and block coalescing \cite{strati2024d,chen2024kvdirect,li2025flowkv}. However, these mechanisms often pursue conflicting objectives. Many achieve speedups by increasing communication concurrency to hide latency, yet none coordinates this concurrency across communication types. As a result, even with these optimizations, collective communication and KV-cache transfers often interfere with one another, which instead amplifies contention.

\section{Methodology}\label{sec:Methodology}

\subsection{Problem Formulation}\label{sec:Methodology:formulation}

\textbf{Simplified Abstraction.} Without loss of generality, we consider a single prefill request processing through $L$ sequential Transformer layers. The communication workload for each layer can be generally abstracted as Multi-stage Flow (MsFlow). An MsFlow consists of three temporally dependent stages, where each stage constitutes a set of flow (or coflow) governed by the layer's lifecycle, which is detailed below:
\begin{enumerate}[label=$\bullet$,leftmargin=*,itemsep=0.5pt]
\item \textbf{Stage~1: Initialization.} Involves KV-cache reuse to transfer prerequisite states. This loosely coupled flow often overlaps with prior layer's computation to hide latency.
\item \textbf{Stage~2: Execution.} Consists of collective communication (e.g., alltoall) that strictly blocks the subsequent computation.
\item \textbf{Stage~3: Completion.} Handles the P2D transfer of results to decoding workers. This dictates the final token availability without blocking the current prefill computation.
\end{enumerate}

For a single request, MsFlows collectively share one TTFT deadline and may partially overlap subject to layer dependency (e.g., Stage~2 of layer $l$ must complete before computing layer $l+1$)

\textbf{Contention dynamics.} Contention arises when concurrent MsFlows compete for shared network bandwidth. Recall Fig~\ref{fig:bg:contention}, intra-request contention occurs between MsFlows of different layers within the same request, while inter-request contention arises between MsFlows of distinct requests. Both cases simply reduce to contention between heterogeneous stages.

\textbf{Overall Objective. } The goal is to meet end-to-end TTFT deadline (SLO attainment) rather than optimizing individual flow latencies. Specifically, we must orchestrate the dependent MsFlows to ensure that all $L$ MsFlows are completed within the deadline constraint. To achieve this, an ideal scheduler must balance two complementary objectives tailored to the specific semantics of MsFlow stages:
\begin{enumerate}[label=(\roman*),leftmargin=*,itemsep=0.5pt]
    \item \textbf{Minimizing Non-overlapped communication Latency.} For early stages (1\&2), completing too late directly prolongs the current layer's execution time, postponing the release of the subsequent layer.
    \item \textbf{Minimizing Earliness.} For the final Stage~3, completing too early yields no gain for meeting deadline but exacerbates potential network contention.
\end{enumerate}

\subsection{Key Observation and Scheduling Principle}
\begin{figure}[t!]
    \centering

    \includegraphics[width=\linewidth]{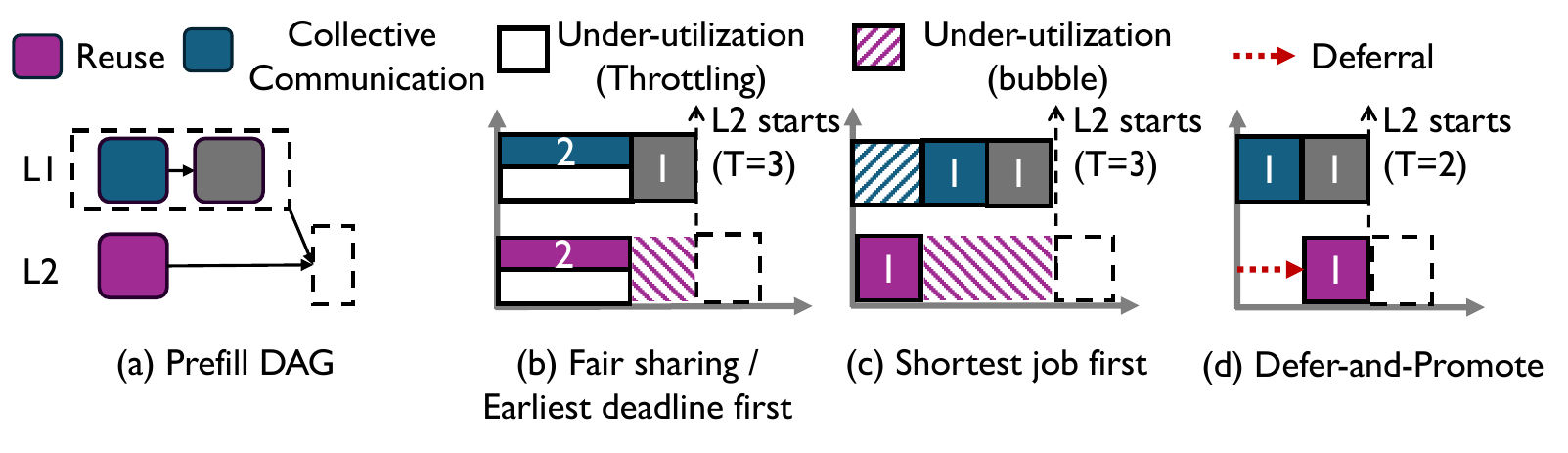}
    \caption{\textbf{Comparison of scheduling policies for intra-request contention (ingress).}
    (a) Ingress port contention during Layer-1 execution.
    (b)-(c) FS/SJF/EDF delays the start of Layer-2 ($T=3$).
    (d) The \textit{Defer-and-Promote} strategy advances the Layer-2 start time to $T=2$ (-33\%).}
    \label{fig:meth:intraA}

    \includegraphics[width=\linewidth]{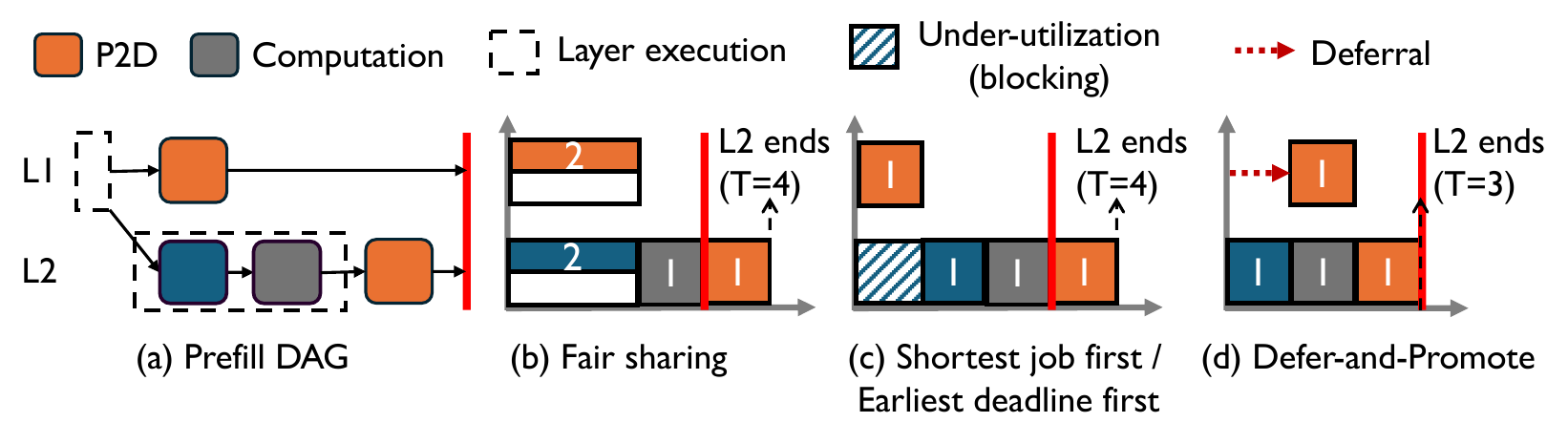}
    \caption{\textbf{Comparison of scheduling policies for intra-request contention (egress).}
    (a) Egress port contention during Layer-2 execution.
    (b)-(c) FS/SJF/EDF delays the end of Layer-2 ($T=4$).
    (d) The \textit{Defer-and-Promote} strategy reduces the Layer-2 finish time to $T=3$ (-25\%).}
    \label{fig:meth:intraB}
\end{figure}

\begin{figure}[t!]
    \centering

    \captionof{table}{An inter-request contention example.}
    \label{tab:InterReqExample}

    \fontsize{9}{11}\selectfont
    \begin{tabular}{c|c|c|c|c}
        \toprule
        \textbf{Req ID} & \textbf{Flow ID} & \textbf{Flow Size} & \textbf{Remain time} & \textbf{Deadline} \\
        \midrule
        1   & A & 2 & 9 & 18 \\
        2,3 & B & 4 & 6 & 12 \\
        4   & C & 3 & 0 & 7  \\
        \bottomrule
    \end{tabular}

    \begin{subfigure}{0.48\columnwidth}
        \centering
        \includegraphics[width=\linewidth]{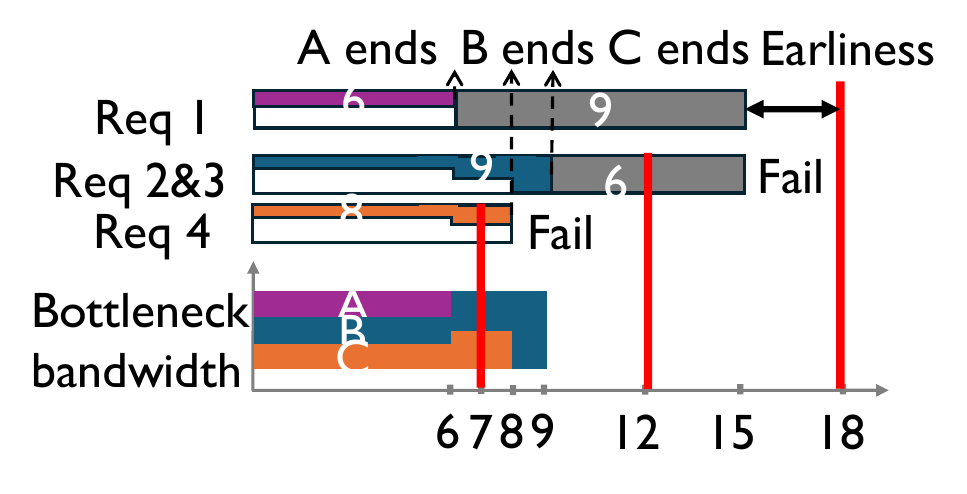}
        \caption{Fair Share}
        \label{fig:bg:expfs}
    \end{subfigure}
    \hfill
    \begin{subfigure}{0.48\columnwidth}
        \centering
        \includegraphics[width=\linewidth]{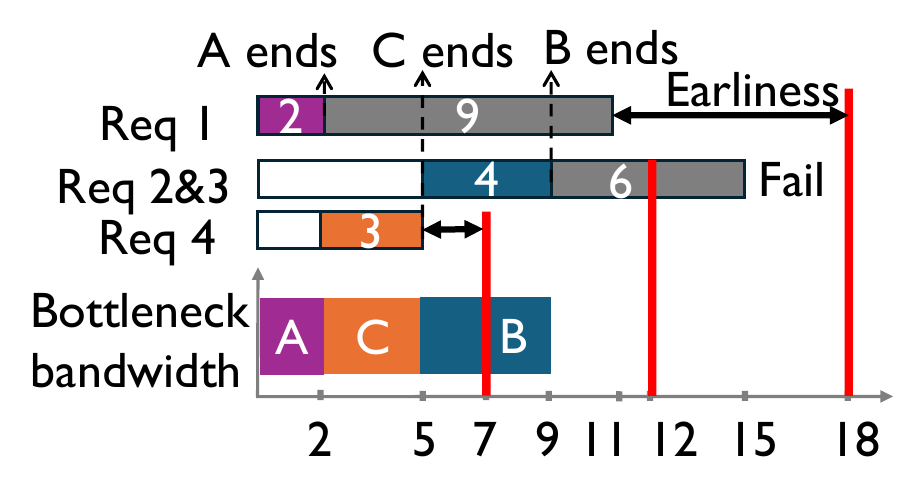}
        \caption{Shortest Job First}
        \label{fig:bg:expsjf}
    \end{subfigure}

    \par\medskip

    \begin{subfigure}{0.48\columnwidth}
        \centering
        \includegraphics[width=\linewidth]{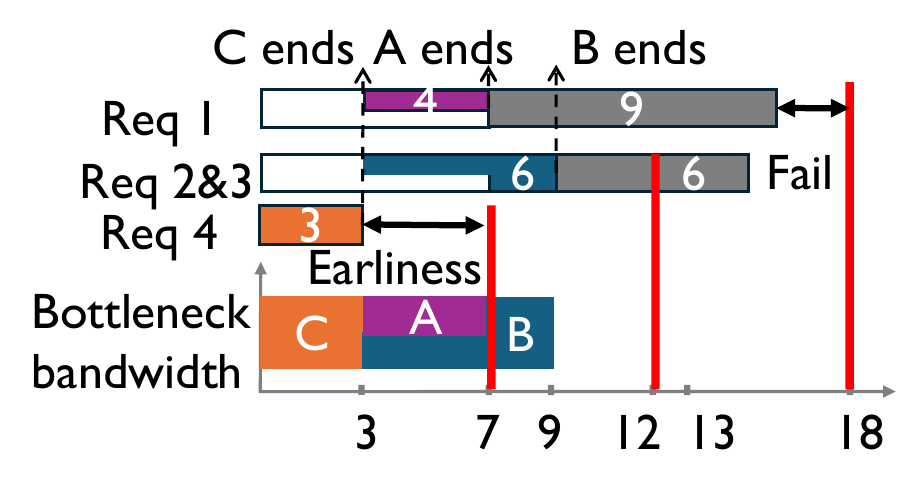}
        \caption{Earliest Deadline First}
        \label{fig:bg:expedf}
    \end{subfigure}
    \hfill
    \begin{subfigure}{0.48\columnwidth}
        \centering
        \includegraphics[width=\linewidth]{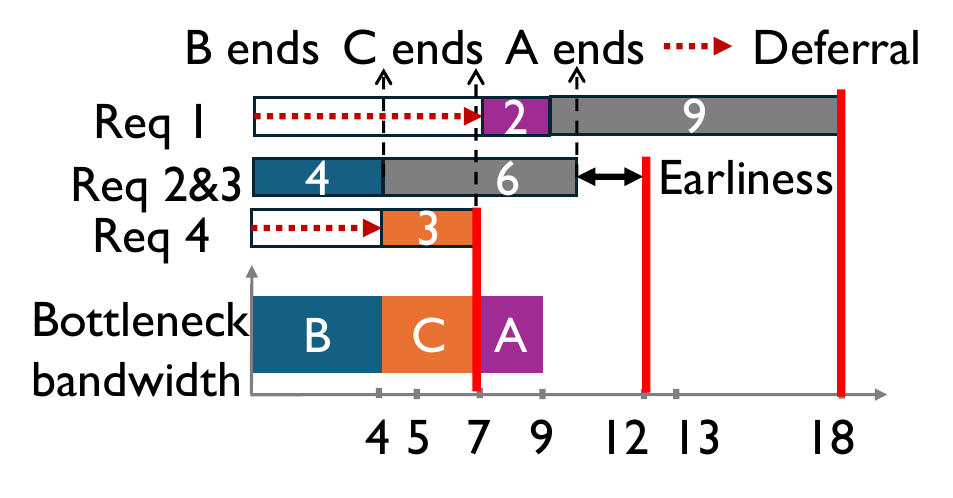}
        \caption{Defer-and-Promote}
        \label{fig:bg:expopt}
    \end{subfigure}

    \caption{\textbf{Comparison of scheduling policies for inter-request contention.}
    The red vertical lines denote hard deadlines, and gray bars are subsequent durations.
    (a) Fair Share and (b) SJF fail to protect the urgent Flow-B due to lack of deadline awareness,
    while (c) EDF completes flows with explicit but loose deadlines unnecessarily early.
    (d) \DAP strategically defers non-urgent flows to minimize earliness,
    and promotes them only when necessary to achieve \textit{just-in-time completion}.}
    \label{fig:meth:inter}

    \vspace{-2em}
\end{figure}

Achieving optimality is theoretically intractable~\cite{bettati1990algorithms,bettati1992end}. To derive an effective heuristic, we first look into the execution dynamics of prefill. Our analysis reveals a structural property of request deadlines that motivates scheduling principle presented in this section.

\parab{Key observation.}
In the prefill stage, \emph{request deadlines progressively materialize into flow-level deadlines as the prefill pipeline advances.}

Recalling Sec.~\ref{sec:Methodology:formulation}, for the initial stages (1\&2), the flow-level deadline is implicit, since it jointly determined by TTFT and duration of downstream tasks. As execution reaches the final stage, this constraint materializes: the global TTFT directly dictates the token return time, becoming an explicit flow-level bound. We borrow the concept of laxity (defined as the remaining budget before triggering deadline violation) from real-time systems to represent flow-level urgency. The distinct laxity reveals a scheduling opportunity: we can safely defer last-stage flows (possessing deterministic laxity) to yield bandwidth for early-stage flows whose laxity is uncertain.

\parab{Scheduling principle: \DAP.} Leveraging the observation above, our key principle is to \emph{defer non-urgent transfer until the latest safe moment, promoting priority only as necessary}, so as to minimize interference while ensuring timely completion. This principle translates into two concrete rules:
\begin{itemize}[leftmargin=*]
    \item \textbf{Principle \#1 (Last stage flows with Explicit-deadline)} are initially deferred to yield bandwidth but gradually promoted to guarantee compliance.
    \item \textbf{Principle \#2 (Early stage flows with implicit-deadline)} are deferred based on relative laxity and promoted as the laxity diminishes to mitigate the violation risks.
\end{itemize}

\parab{Why this works.}
By deferring non-urgent flows and selectively promoting them when necessary, the scheduler balances short-term efficiency with long-term deadline guarantees.
This strategy improves TTFT SLO attainment along two axes:
(i) minimizing the non-overlapping communication latency to shorten the prefill makespan, and
(ii) regulating earliness to prioritize tighter requests under contention.

To see why \DAP benefits the prefill DAG makespan, consider the  consider the contention scenarios in Fig~\ref{fig:meth:intraA}-(a) (ingress) and Fig~\ref{fig:meth:intraB}-(a) (egress). Fair Sharing indiscriminately dilutes Stage~2 throughput; SJF (Fig.~\ref{fig:meth:intraA}-(c), \ref{fig:meth:intraB}-(c)) preferentially prioritizes KV-movement flows (flow size typically smaller); and EDF over-prioritizes Stage~3 due to explicit deadlines (Fig.~\ref{fig:meth:intraA}-(c)) and degenerates into Fair Sharing when deadlines are implicit (Fig.~\ref{fig:meth:intraB}-(b)). Consequently, all three policies inflate the non-overlapped communication latency and stall computation. Defer-and-Promote resolves this by strategically deferring Stage~3 (higher laxity) to protect Stage~2, which demonstrating a significant reduction in makespan (Fig.~\ref{fig:meth:intraA}-(d), \ref{fig:meth:intraB}-(d)).

\DAP maximizes attainment under inter-request contention by strictly prioritizing truly urgent requests. Consider Table~\ref{tab:InterReqExample} as an example, where each request has an end-to-end TTFT constraint and implicit remaining time determined by downstream tasks. As illustrated in Fig.~\ref{fig:meth:inter}, three flows with varying sizes and deadlines contend for a bottleneck link. Unlike standard policies (e.g., FS, SJF, EDF) that often misjudge urgency, \DAP prioritizes Collective Communication, as deferral would immediately stall the execution pipeline. It strategically defers other traffic: Stage~1 flows are promoted only when they threaten to stall the pipeline, while last-stage flows are delayed until their explicit deadlines.

\section{Design}

\subsection{Design Challenges}

While the \DAP{} principle is conceptually simple, realizing it in LLM serving is challenging in practice. \sys{} must make priority assignments \emph{in the dark}, steering flows without precise laxity through a \emph{discrete multi level queue} (e.g., finite switch queues). Specifically, we identify the following two challenges.

\parab{C-1: Scheduling flows with explicit deadlines without excessive prioritization.}
Even when flows carry explicit deadlines (e.g., P2D transfers), determining when to switch from defer to promote is non-trivial under discrete priority levels. Promoting too early wastes scarce bottleneck bandwidth, while promoting too late risks deadline misses. The challenge lies in striking the right balance for just-in-time completion given finite priority classes.

\parab{C-2: Scheduling implicit-deadline flows without precise laxity.}
The absence of precise laxity traps the scheduler in an error-prone dilemma. On one hand, underestimating laxity incurs unintended execution stalls, as mistakenly deferring Stage~2 flows immediately idles the GPU and inflates request latency. On the other hand, overestimating laxity results in priority inversion, squandering scarce bandwidth on deferrable work and starving truly urgent requests.

\begin{figure}[t!]
    \centering
    \includegraphics[width=\columnwidth]{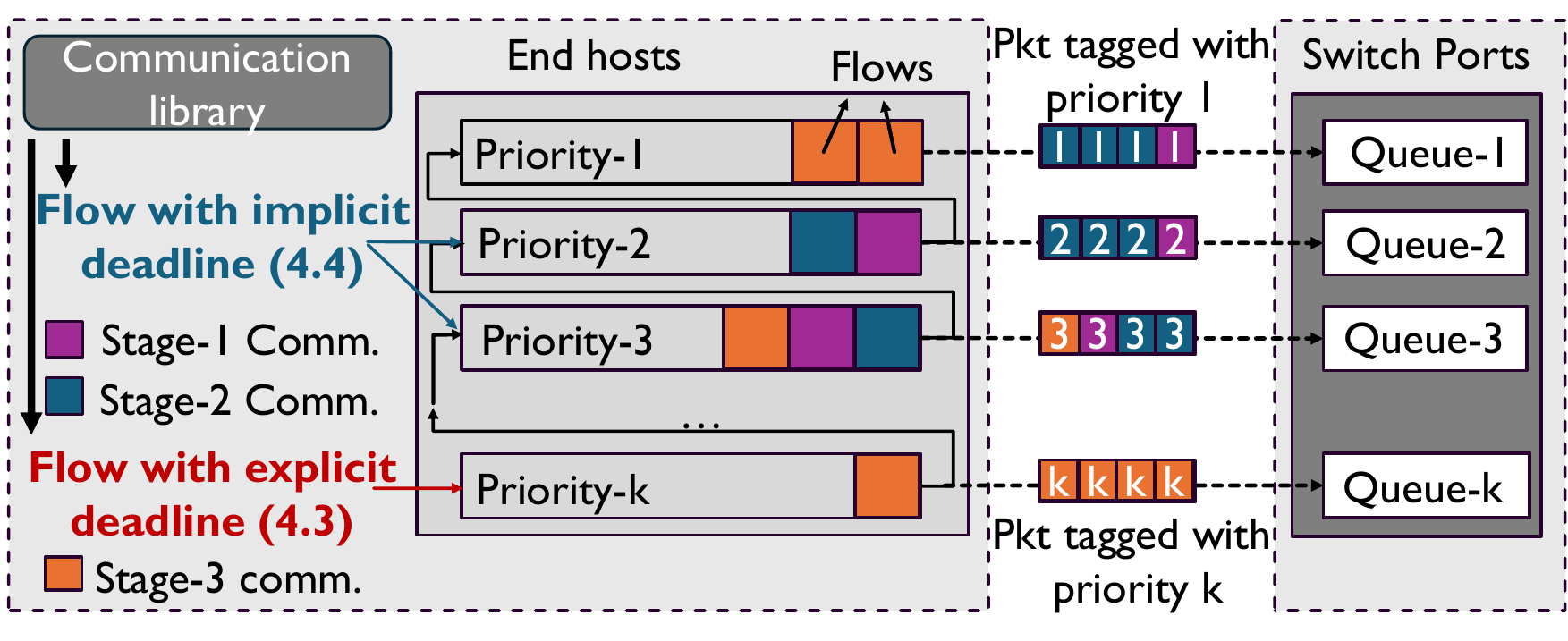}
    \caption{\textbf{Design overview.}
    }
    \label{fig:design:designoverview}
\end{figure}

\subsection{Overview} Figure~\ref{fig:design:designoverview} illustrates \sys{}'s core abstraction: the \emph{Reverse Multi-Level Queue (RMLQ)}. To realize the \DAP{} principle, RMLQ inverts the classic MLFQ paradigm (which demotes flows over time); instead, it initializes flows in low-priority queues to enforce deferral, promoting them strictly when diminishing laxity necessitates immediate execution.

At a high level, \sys{} schedules all multi-stage communication through a shared RMLQ substrate, while applying separate rules (initial priority and promotion pace) for different stages. For last stage flows with explicit deadlines, \sys employs Minimal Link Utilization (MLU), promoting a flow only when the remaining bandwidth is just insufficient to meet its deadline, thereby avoiding premature over-prioritization. For early stage flows with implicit deadline, \sys speculate relative urgency via Relative Layer Index (RLI). Flows begin at lower tiers climb up when it threaten to stall computation. Finally, to resolve potential cross-stage contention, \sys{} reserves the highest priority level for deadline-explicit flows to avoid violations, and prioritizes deadline-implicit flows in the remaining levels to opportunistically exploit available bandwidth.

The rest of this section is organized as follows. \S\ref{design:explicit} introduces scheduling flows with explicit deadlines. \S\ref{design:implicit} describes the strategy for implicit deadlines. Finally, \S\ref{sec:design:PET} combines these components to present the complete RMLQ arbitration logic.

\subsection{Scheduling with Explicit Deadline}\label{design:explicit}

To address C-1, we need to ensure the last-stage communication task is judiciously deferred—neither excessively prioritized nor starved—while strictly guaranteeing the request deadline. The challenge lies in mapping continuous urgency levels derived from these deadlines onto $K$ discrete priority queues. Consequently, the scheduler needs a mechanism to approximate continuous deadline-derived urgency using coarse-grained priority levels, which requires determining how to quantify urgency, when to trigger promotion, and at what granularity to apply scheduling decisions (e.g., per-packet or per-layer).

Specifically, consider an $L$-layer model serving a batch of request. At each layer $\ell$, the P2D stage emits a set of flows. Flows within the same request $r$ inherit its TTFT deadline $D_r$, while flows from different requests are scheduled independently. We assume there are K priority queues $P_i(1\leq i \leq K)$ where $P_1$ has the highest priority. We denote the threshold for promoting the priority from $P_{j}$ to $P_{j-1}$ as $\tau_{j}$ ($2 \le j \le K$). we define $\tau_K = +\infty$ ensuring that even the extreme loose flows are captured in the lowest-priority queue.

\parab{Urgency metric.} We define \emph{Minimal Link Utilization (MLU)}: $\mathrm{MLU}_i(t)=\frac{\textit{Size}_{rem}(t)}{\textit{Time}_{rem}(t)\cdot B \cdot (1-\rho)}$, where $B(1-\rho)$ represents the effective bandwidth after accounting traffic load $\rho$. This metric represents the minimal share of the residual link capacity required to satisfy the deadline. Crucially, $\mathrm{MLU}_i(t) > 1$ signifies an infeasible overload state; values approaching $1$ signal critical urgency requiring exclusive service, while lower values imply sufficient slack for deferral. By contrast, purely deadline-based metrics (e.g., EDF) are ill-suited since they suffer from the domino effect under overload \cite{buttazzo1995value} and further degrade when constrained to discrete priority levels \cite{buttazzo2005rate}.

\parab{Optimizing promotion threshold.} Deriving the globally optimal thresholds that minimize deadline miss rates is computationally intractable (NP-hard). Therefore, we seek a robust, practical approximation by minimizing the relative quantization error introduced when mapping continuous urgency to discrete levels. When a flow with urgency $v \in [U_{min},U_{max}]$ is mapped to a discrete priority level $\tau_k$, it incurs a relative error of $\epsilon \approx \frac{|v - \tau_k|}{v}$. Our intuition here is to minimize the worst-case relative error across all threshold. $\prod_{k=1}^{K} r_k = \frac{u_2}{u_1} \cdot \frac{u_3}{u_2} \cdot \dots \cdot \frac{u_K}{u_{K-1}} = \frac{u_K}{u_1} = \frac{U_{\max}}{U_{\min}}$. Mathematically, this minimum is achieved if and only if all ratios are equal ($r_k \equiv r$). This necessitates a geometric progression for threshold generation (where $r =(U_{\max}/U_{\min})^{\frac{1}{K-1}}$). Since the precise bounds $U_{\max}$ and $U_{\min}$ are often unknown or dynamic in practice, we approximate this geometric spacing by configuring the promotion threshold as $Q_i = E^{-i}\cdot U$ ($1\le i\le K-1$), where parameters $E$ and $U$ are set empirically (e.g., $E=4, U=0.5$) to yield robust performance.

\parab{Promotion behavior.}
Fine-grained promotion decision may adversely introduce practical concerns in network scheduling. In particular, if promotion is applied at a fine granularity (e.g., per packet), a single message may be fragmented across multiple priority queues. This can lead to packet reordering, where later packets overtake earlier ones, triggering costly transport-layer recovery and reducing effective throughput. To avoid this issue, we restrict promotion to layer boundaries, ensuring priority atomicity at the message level. Crucially, this coarser granularity does not compromise scheduling expressiveness, as the substantial depth of modern LLMs ($L>K$, e.g, L=64 and K=8/16) provides ample room for priority adjustments. Additionally, we enforce a monotonic promotion policy (i.e., only allowing promotion). This prevents priority oscillation and reinforces the \DAP{} principle, ensuring that flows are conservatively maintained in lower tiers until urgency strictly necessitates promotion.

\subsection{Scheduling with Implicit Deadlines} \label{design:implicit}

To address the lack of precise laxity under implicit deadlines (Challenge C-2), \sys exploits structural determinism to infer relative urgency within requests, and resolves residual contention across requests using robust arbitration mechanisms.

\subsubsection{Intra-request Scheduling with Structural Determinism}\label{design:intrareq}

While precise laxity is elusive due to dynamic scheduling shifts, the relative laxity within request is deterministic: flows blocking immediate execution strictly more urgent than lookahead transfers that enable future progress.

We quantify this using the \emph{Relative Layer Index}, defined as $\text{RLI} = L_{\text{target}} - L_{\text{curr}}$, where $L_{\text{curr}}$ is the index of the currently executing layer and $L_{\text{target}}$ is the target layer index where the data is consumed. A smaller RLI indicates a tighter safe deferral window. \sys{} resolves contention by strictly prioritizing flows with the smallest RLI. We apply it to resolve two representative contention cases detailed below.

\parab{Case I: Collective vs. KV Reuse.} When $\text{collective}^{i}$ competes with $\text{reuse}^{j}$ ($j > i$), the collective targets the current layer ($L_{\text{target}} = i$), yielding $\text{RLI}(\text{coll}) = 0$. In contrast, the reuse targets layer $j$ with $\text{RLI}(\text{reuse}) > 0$. We therefore enforce $\text{collective}^{i} \succ \text{reuse}^{j}$ to avoid execution stalls.

\parab{Case II: Inter-KV Reuse Contention.} Under multi-layer lookahead, flows $\text{reuse}^{j}$ and $\text{reuse}^{k}$ ($L_{\text{curr}} < j < k$) often contend. Comparing their urgency, layer $j$ represents a nearer deadline, implying $\text{RLI}(j) < \text{RLI}(k)$. Prioritizing the smaller RLI effectively delays the onset of the earliest execution stall. Thus, we enforce $\text{reuse}^{j} \succ \text{reuse}^{k}$.

In summary, \sys{} unifies these decisions by strictly prioritizing flows with the smallest RLI. Theorem 1 formalizes the optimality of this structural ordering.

\parab{Theorem 1.}  \emph{Assuming an ideal computation model without preemption overhead, the prefill makespan is minimized by a schedule that strictly prioritizes flows with the smallest RLI.}

\emph{Proof sketch.} Intuitively, prioritizing a high-RLI flow during a stall wastes the opportunity to overlap it with subsequent computation. We prove this via an exchange argument: shifting any flow with a larger RLI out of a stall interval into a later overlapped region strictly reduces the current blocking duration without violating future dependencies. We provide the detailed proof in Appendix.

\subsubsection{Robust Inter-request Scheduling}\label{design:interreq}

With intra-request contention resolved by RLI (\S\ref{design:intrareq}), the remaining challenge is breaking ties \emph{across requests} when competing flows have similar RLI and simultaneously threaten execution. The scheduler must protect requests with high SLO violation risks under imprecise laxity. However, synchronous batch execution exacerbates the impact of scheduling pitfalls, allowing errors to propagate to all batched peers. Our intuition is not to recover exact laxity, but to make scheduling decisions \emph{robust to estimation errors}, prioritizing requests/batches where delays would most severely hurt SLO attainment.

\parab{Scheduling pitfalls under imprecise laxity.}
We identify two representative failure cases arising from misjudging laxity:
\begin{itemize}[leftmargin=*]
    \item \textbf{The Piggyback effect.} In batches with mixed deadlines, a small number of extremely tight requests can dominate the inferred batch-level urgency. Consequently, loose requests in the same batch inherit elevated priority, preempting other batches that are globally more urgent.
    \item \textbf{The Black Hole effect.} Proximity to a deadline does not guarantee feasibility. Under overload, a batch may appear urgent while carrying a workload that exceeds available bandwidth. Prioritizing such batches wastes resources on inevitable failures, blocking other viable batches.
\end{itemize}

We define the \emph{Robust Effective Deadline} ($RED$) to quantify urgency while counteracting the Piggyback effect. We partition a batch into tight and loose sub-batches at the maximal deadline gap. With $D^T_{\text{min}}$ and $D^{Lo}_{\text{min}}$ representing the minimum deadline of each sub-batch, we define $RED = f \cdot D^T_{\text{min}} + (1-f) \cdot D^{Lo}_{\text{min}}$, where $f$ is the proportion of tight requests. When tight requests are rare (small $f$), $RED$ naturally shifts towards the loose deadline ($D^{Lo}_{\min}$), preventing outliers from hijacking the batch priority.

Complementing RED-based ordering, we employ overload control to mitigate the Black Hole effect. The intuition is to prevent doomed or excessively heavy requests from blocking viable ones. We perform a worst-case feasibility check by accumulating estimated computation and communication delays. We iteratively prune the requests contributing the most to the cumulative delay to restore feasibility for the remainder of the batch.

The full algorithm, detailed in the Appendix, synthesizes these strategies. Triggered upon batch arrival and departure, it outputs two key results: (1) an ordered sequence $\sigma$ derived from RED, and (2) a set of pruned requests $\mathcal{H}$ identified as infeasible. We deliberately avoid fine-grained per-layer updates to prevent the scheduler from over-reacting to transient load estimation jitter.

\subsection{Putting Everything Together}\label{sec:design:PET}

\sys schedules all traffic through a unified RMLQ substrate. We classify flows according to stage feature and apply distinct rules for initial priority assignment and subsequent promotion.

\begin{itemize}[leftmargin=*]
    \item For last stage flows with explicit deadline, we calculate the MLU defined in\S\ref{design:explicit} to determine the initial priority level. For promotion, \sys{} updates priority at layer boundaries while the request is computing, and switches to periodic updates at fixed intervals after computation finishes.
    \item For early stage flows with implicit deadline, priority assignment relies on the RLI defined in\S\ref{design:intrareq}. Stage~1 flows are initialized based on RLI and promoted incrementally at layer boundaries to align with computation progress. Stage~2 flows directly enter the high priority queue.
\end{itemize}

\parab{Arbitration.} \sys strictly reserves the highest priority for the most urgent last-stage flows. Within the remaining priority levels, \sys adopts a hierarchical policy: it grants precedence to early-stage flows over last-stage flows to opportunistically exploit available bandwidth; to break ties among early-stage flows sharing the same RLI, \sys adheres to the rank determined by the inter-request scheduling (\S\ref{design:interreq}).

\section{Implementation}

We implement \sys{} in approximately 10,000 lines of code. \sys employ two-tier control plane composed of local daemons and a centralized coordinator. Local daemons maintain scheduling context and enforce scheduling decisions, while reporting request-level statistics to the coordinator upon scheduling and completion events. The coordinator operates at request granularity, aggregating these summaries to perform inter-node tie-breaking, and disseminates the resulting decisions back to local daemons for enforcement.

\parab{System Integration and Compatibility.}
We implement \sys as a pluggable module that integrates transparently with NCCL and Mooncake. Each node runs a lightweight local daemon that interfaces with runtimes via lock-free MPSC queues over shared memory, enabling low-latency intra-node signaling. By intercepting low-level network primitives (e.g., RDMA Verbs), \sys redirects work requests from the library's proxy threads to the local daemon, which orchestrates task execution according to our scheduling logic. The interaction between the runtime and the scheduler is governed by three standardized primitives: (1) \texttt{submit} registers tasks with metadata, (2) \texttt{permit} gates transmission by granting specific priorities, and (3) \texttt{completion} updates the internal state upon hardware signals. Requests can be pruned by modifying their metadata to suppress communication.

\parab{Priority enforcement.} Commodity hardware typically exposes a limited number of priority classes ($K \ll L$), making direct mapping infeasible. In practice, collective traffic is typically confined to a regional serving unit, whereas KV-cache traffic spans the network across racks. We employ a hybrid priority queue tailored for such traffic. Specifically, \sys enforces strict priority at the host via software queues (arbitration logic in \S\ref{sec:design:PET}), while leveraging DSCP-based hardware priorities to isolate KV-cache traffic within the network fabric. For flows with explicit deadlines, we calculate MLU threshold derived from $K$ and the current network load $\rho$, and assign DSCP tags according to this threshold. For flows with implicit deadlines, we map their RLI to the physical queue range $[0, K-1]$ by capping the RLI value at $K-1$.

\section{Evaluation}
\label{sec:evaluation}

\subsection{Experiments setup}

\parab{Testbed setup.}
The testbed consists of 8 servers, each equipped with 4 NVIDIA GeForce RTX 3090 GPUs, 40 CPU cores (Intel(R) Xeon(R) Gold 5218R CPU @ 2.10GHz), 128GB RAM, and two ConnectX-5 100Gbps NIC. The servers are connected via NVIDIA SN2700 Ethernet switches. Within each server, the GPUs communicate via PCIe Gen 3.0. All servers run Ubuntu 22.04 with CUDA 12.8, PyTorch 2.8.0, and NCCL 2.28. we configure a prefill cluster comprising 16 GPUs, while decode workers are provisioned separately. For simplicity, we limit the decode output length to one token, since it does not affect TTFT.

\parab{Simulation setup.}
Our simulator is built upon Vidur~\cite{agrawal2024vidur} and the flow-level simulator flowsim~\cite{FlowSim2024}. We extend Vidur to support the Mixtral model with expert and sequence parallelism, as well as PD disaggregation with KV-cache reuse. The simulation workflow begins with offline profiling of operator latencies via Vidur profiler. At runtime, we employ a unified event-driven mechanism: Vidur simulates the LLM serving behavior to generate computation and communication tasks, while flowsim parser the communication task and trigger the underlying network events. Crucially, both computation events and network events are processed within a single event queue to ensure correctness. By default, we simulate a 256-server cluster with a 1:1 prefill-decode worker ratio, managed by a KV-cache aware scheduler similar to Dynamo~\cite{NVIDIA_Dynamo}.  Each server hosts eight GPUs interconnected via NVSwitch (900 GB/s) and eight NICs, connected via a 1:1 fat-tree network with link bandwidth 200Gbps. The latency profiles are calibrated based on NVIDIA A100 GPU.

\parab{Metrics.}
We primarily evaluate Time-to-First-Token (TTFT) and its Service Level Objective (SLO) attainment rate. Following the methodology in~\cite{patke2024queue,hongsola,stojkovic2025dynamollm}, we define the SLO threshold as $3\times$ the TTFT measured under low-load conditions by default. We also evaluate microbenchmarks such as the collective completion time (CCT) and request earliness to help interpret the observed performance gains.

\subsection{Testbed Experiments}
\label{sec:evaluation:testbed}

\begin{figure}[t!]
    \centering
    \begin{subfigure}[t]{0.48\linewidth}
        \centering
        \includegraphics[width=\linewidth]{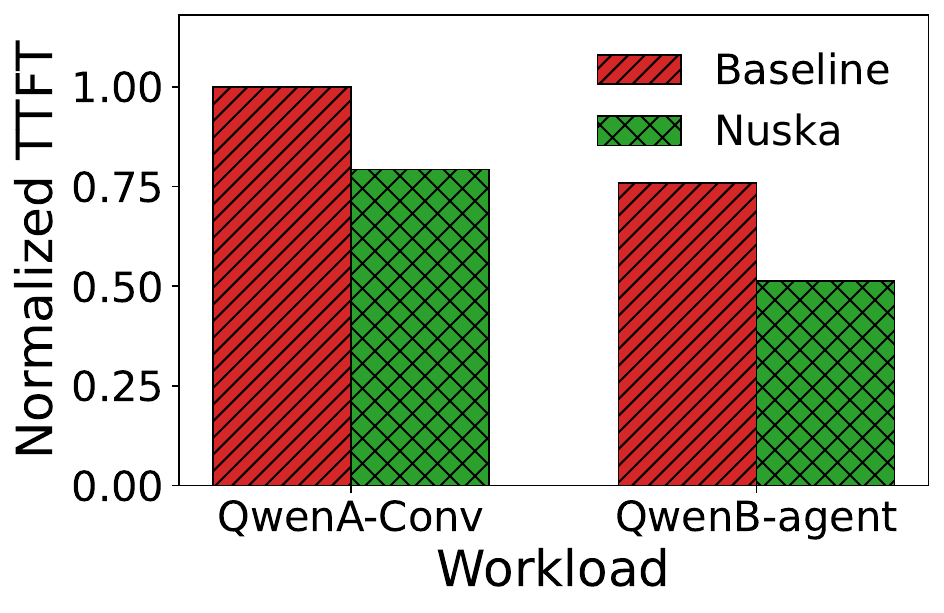}
        \caption{Normalized TTFT.}
        \label{fig:evaluation:testbed:ttft}
    \end{subfigure}
    \hfill
    \begin{subfigure}[t]{0.48\linewidth}
        \centering
        \includegraphics[width=\linewidth]{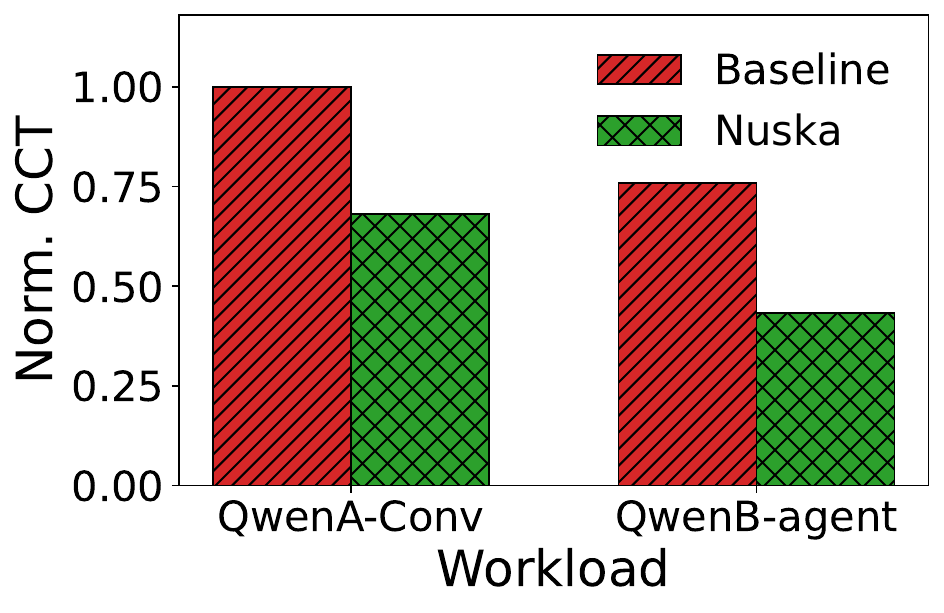}
        \caption{Normalized CCT.}
        \label{fig:evaluation:testbed:cct}
    \end{subfigure}
    \hfill
    \caption{[Testbed] Mixtral-8x7B}
    \label{fig:evaluation:testbed}
\end{figure}

\begin{figure*}[ht!]
    \centering
    \begin{subfigure}[t]{0.9\linewidth}
        \centering
        \includegraphics[width=0.9\linewidth]{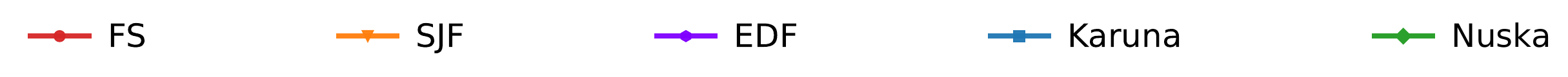}
    \end{subfigure}
    \vspace{-0.5em}
    \begin{subfigure}[b]{0.24\linewidth}
        \centering
        \includegraphics[width=\linewidth]{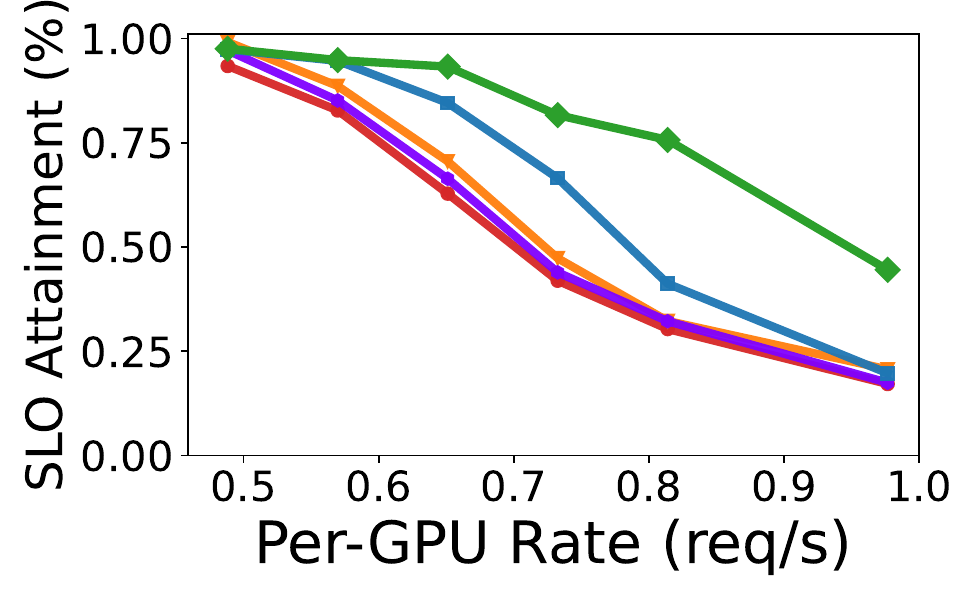}
        \caption{Mixtral-8x22B}
        \label{fig:evaluation:qwenA:mixtral_200}
    \end{subfigure}
    \hfill
    \begin{subfigure}[b]{0.24\linewidth}
        \centering
        \includegraphics[width=\linewidth]{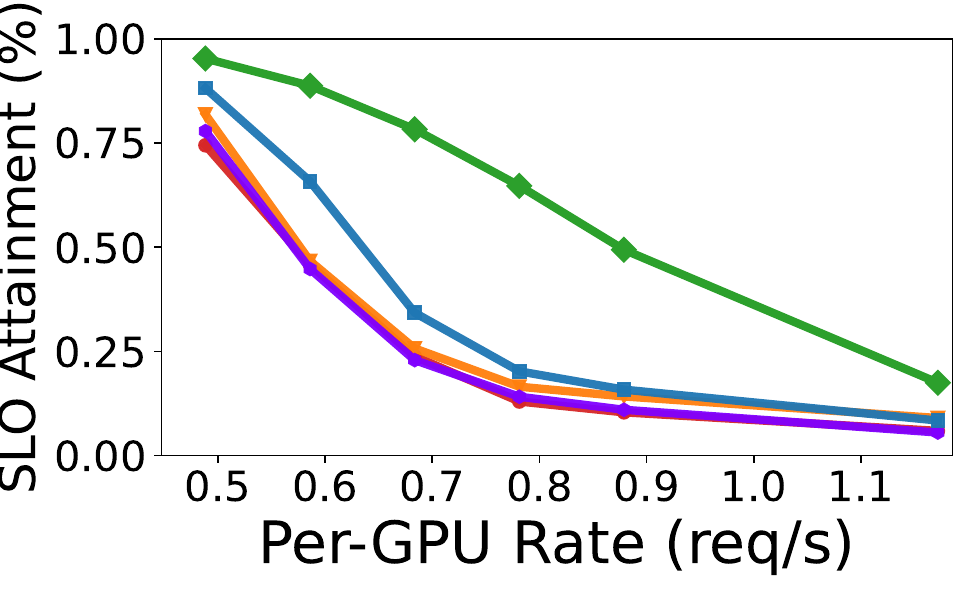}
        \caption{DBRX}
        \label{fig:evaluation:qwenA:dbrx_200}
    \end{subfigure}
    \hfill
    \begin{subfigure}[b]{0.24\linewidth}
        \centering
        \includegraphics[width=\linewidth]{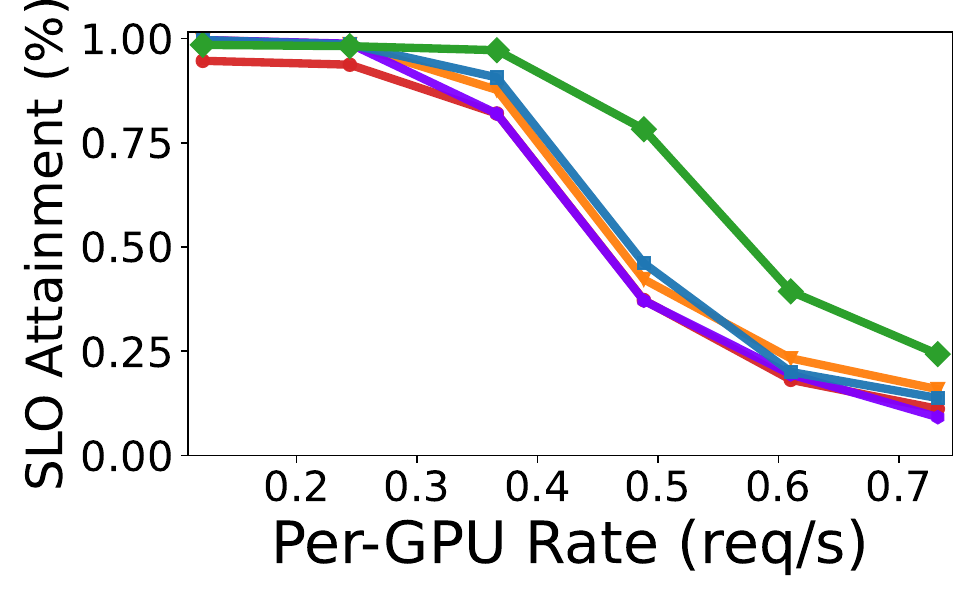}
        \caption{Qwen3-Coder}
        \label{fig:evaluation:qwenA:qwencoder_200}
    \end{subfigure}
    \hfill
    \begin{subfigure}[b]{0.24\linewidth}
        \centering
        \includegraphics[width=\linewidth]{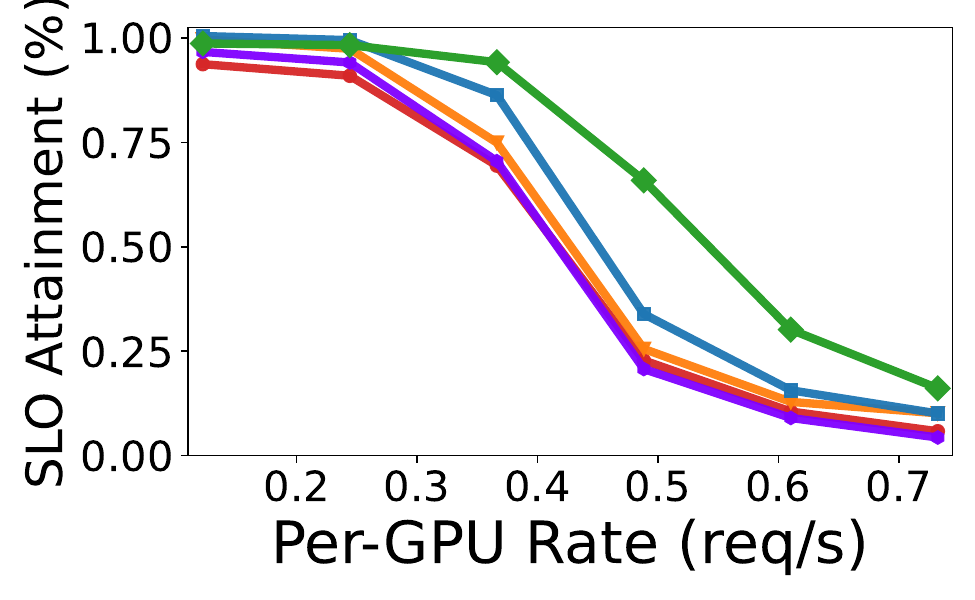}
        \caption{Grok2}
        \label{fig:evaluation:qwenA:grok2_200}
    \end{subfigure}
    \caption{[Simulation] Performance on conversation workload (QwenA)
    }
    \label{fig:evaluation:qwenA}
\end{figure*}

\begin{figure*}[ht!]
    \begin{subfigure}[b]{0.24\linewidth}
        \centering
        \includegraphics[width=\linewidth]{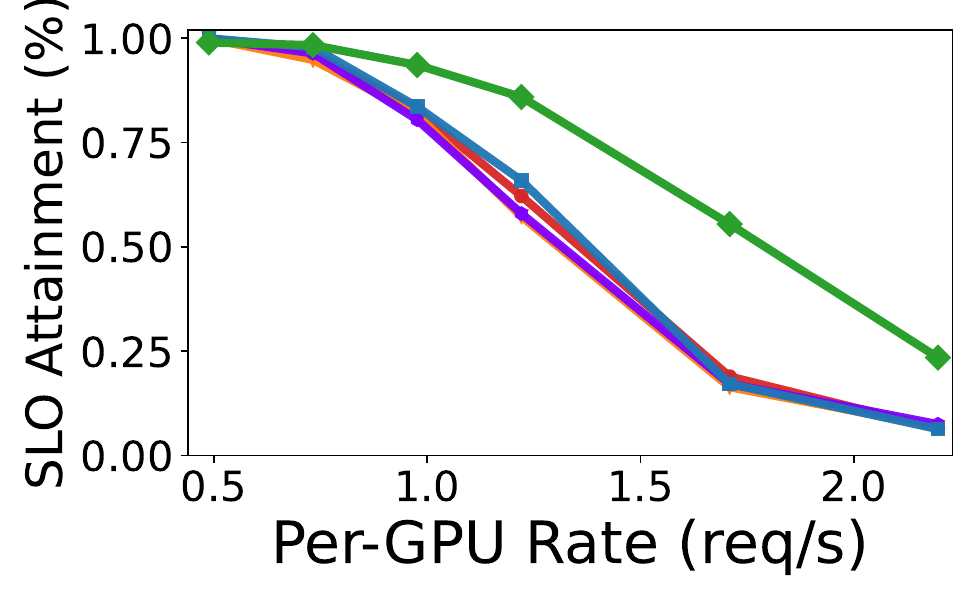}
        \caption{Mixtral-8x22B}
        \label{fig:evaluation:qwenB:mixtral_200}
    \end{subfigure}
    \hfill
    \begin{subfigure}[b]{0.24\linewidth}
        \centering
        \includegraphics[width=\linewidth]{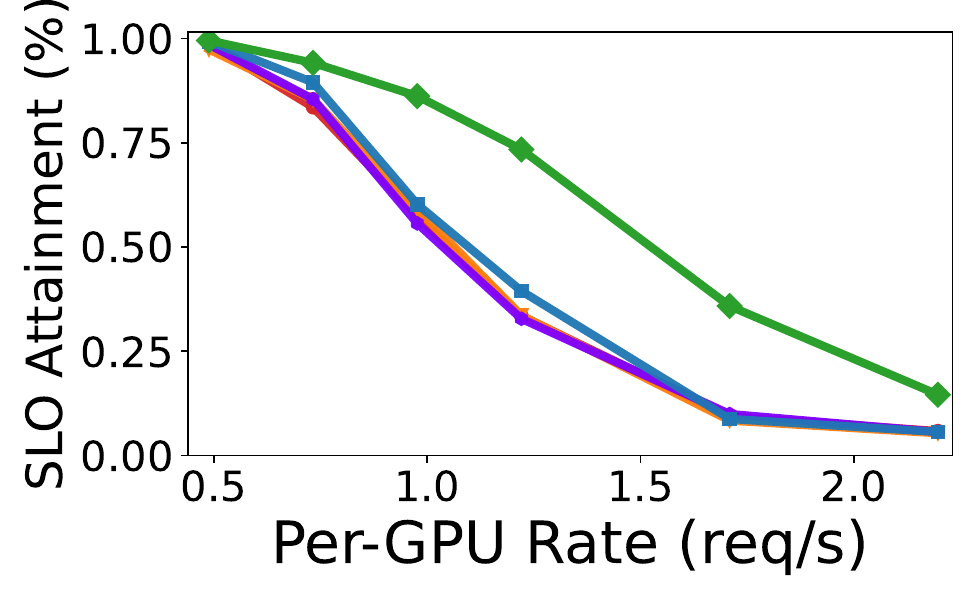}
        \caption{DBRX}
        \label{fig:evaluation:qwenB:DBRX}
    \end{subfigure}
    \hfill
    \begin{subfigure}[b]{0.24\linewidth}
        \centering
        \includegraphics[width=\linewidth]{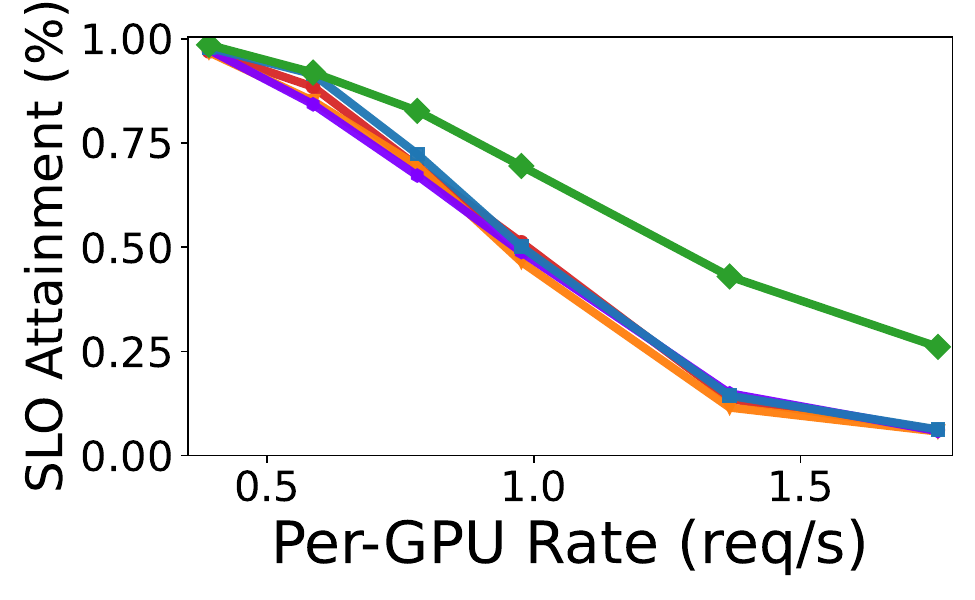}
        \caption{Qwen3-Coder}
        \label{fig:evaluation:qwenB:qwencoder_200}
    \end{subfigure}
    \hfill
    \begin{subfigure}[b]{0.24\linewidth}
        \centering
        \includegraphics[width=\linewidth]{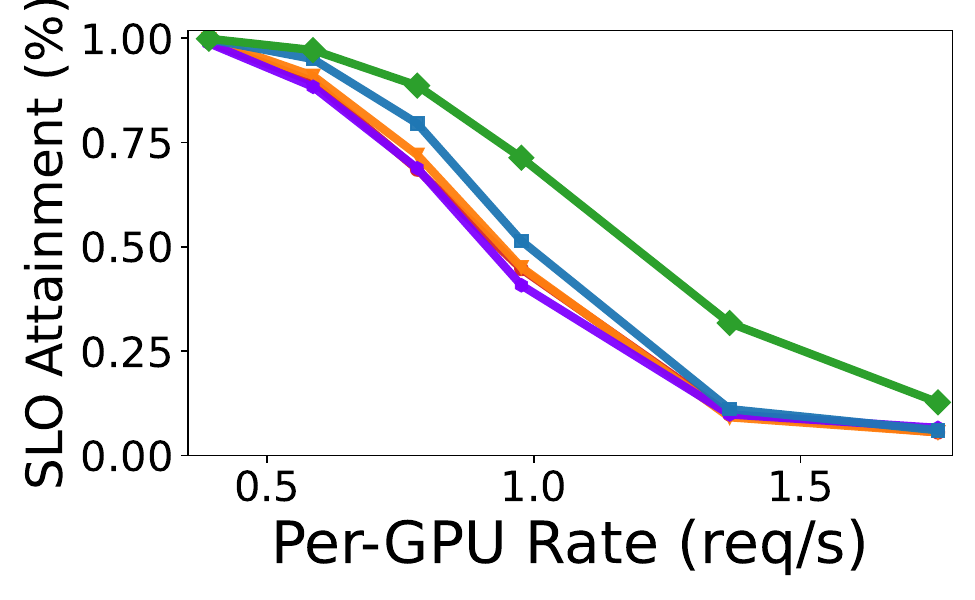}
        \caption{Grok2}
        \label{fig:evaluation:qwenB:grok2_200}
    \end{subfigure}
    \vspace{-0.3cm}
    \caption{[Simulation] Performance on agent workload (QwenB)}
    \label{fig:evaluation:qwenB}
\end{figure*}

\parab{Model and workload.}
In the testbed experiment, We evaluate the Mistral 8x7B~\cite{mixtral7b} MoE model (FP16, top-$k$=2), configured with a tensor parallelism (TP) degree of 1 and expert parallelism (EP) of 8. For the workload, we utilize two production traces derived from Qwen~\cite{qwen_bailian_traces}. QwenA-Conv represents a \textit{conversation workload} with an average sequence length of 2k tokens and a 50\% prompt reuse rate. QwenB-agent represents an \textit{agent workload}, characterized by an average length of 1k tokens and a higher reuse rate of 65\%. Request arrivals follow a Poisson process, where we vary the requests per second (RPS) to modulate system load. Due to testbed capacity limits, we clip the first 512 requests of each trace as a warm-up phase and use the subsequent 1024 requests for evaluation.

Figure~\ref{fig:evaluation:testbed} reports how \sys{} on QwenA-Conv and QwenB-Agent workload. We configure vLLM with 8192 batched tokens and set the per-GPU request rate to one request per second. \sys{} significantly reduces TTFT across both workloads, optimizing the mean TTFT by 20.7\% (1.26×) on QwenA-Conv and 32.3\% (1.48×) on QwenB-Agent. The TTFT reduction is primarily driven by faster completion of computation-blocking collectives. As shown by the all-to-all completion time (CCT), \sys{} shortens stage-2 communication by 31.9\% (1.47×) on QwenA-Conv and 43.1\% (1.76×) on QwenB-Agent. These collectives lie directly on the critical path of prefill execution: delaying them immediately stalls GPU computation and inflates end-to-end latency.

\subsection{Large-Scale Simulations}

In the large-scale simulations, we expand the evaluation scope to cover a diverse spectrum of production-grade architectures and workloads.
\begin{itemize}[leftmargin=*]
    \item We examine a comprehensive set of mainstream MoE models with expert parallelism, ranging from architectures with large but few experts (e.g., Mixtral 8x22B~\cite{mixtral22b} and Grok-2~\cite{grok}, configured with TP=4, EP=8) to models with small but many experts (e.g., DBRX~\cite{dbrx} with TP=2, EP=16, and Qwen3-Coder~\cite{qwen-coder} with TP=1, EP=32). These models are deployed with 32-GPU instance utilizing a hybrid parallelism strategy. We evaluate these MoE models using Qwen conversation and agent traces to represent typical interactive behaviors.
    \item We further assess the effectiveness of \sys{} on dense models with sequence parallelism~\cite{ringattention}. We evaluate the Llama~\cite{llama} model configured with a 1M token context window. In this setup, the model is deployed on 16-GPU instance using TP=4 and SP=4. We pair this model with the Mooncake dataset~\cite{mooncake} to simulate real-world conversation and agent workloads characterized by similar reuse ratios ($\sim40\%$, $\sim65\%$) and extended context lengths ($\sim15k$, $\sim9k$).
\end{itemize}

\parab{Baselines.} We compare \sys{} against four classic flow scheduling policies:
\begin{itemize}[leftmargin=*]
    \item \textbf{Fair Sharing} enforces max-min fairness among concurrent flows by allocating bandwidth equally, regardless of flow sizes or deadlines.
    \item \textbf{Shortest Job First (SJF)} minimizes average flow completion time by strictly prioritizing flows with smaller flow sizes.
    \item \textbf{Earliest Deadline First (EDF)} prioritizes flows with explicit deadlines. Since application-level deadlines do not directly translate to individual network flow deadlines, it applies fair sharing for flows with implicit deadlines.
    \item \textbf{Karuna~\cite{karuna}} allocates the minimum required bandwidth to flows with deadlines to ensure on-time completion, while scheduling the remaining flows using SJF.
\end{itemize}

\begin{figure}[t!]
    \centering
    \begin{subfigure}[b]{0.48\linewidth}
        \centering
        \includegraphics[width=\linewidth]{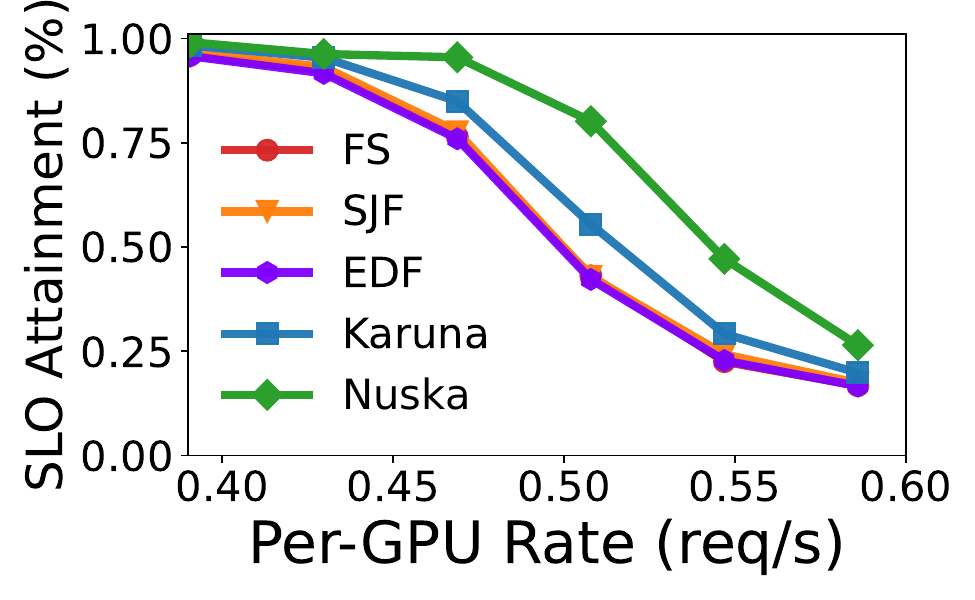}
        \caption{Conversation workload }
        \label{fig:evaluation:sp:conv}
    \end{subfigure}
    \hfill
    \begin{subfigure}[b]{0.48\linewidth}
        \centering
        \includegraphics[width=\linewidth]{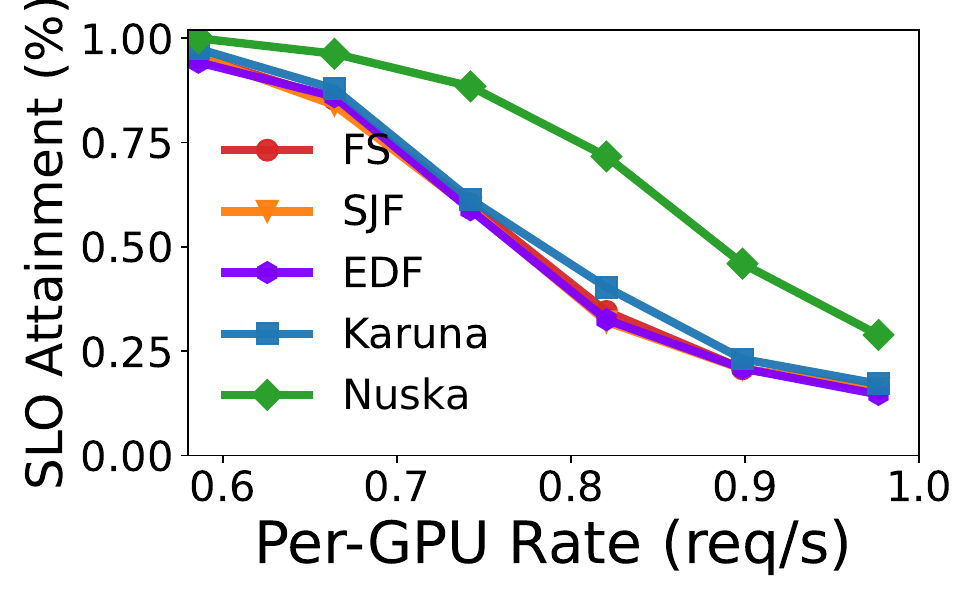}
        \caption{Agent workload}
        \label{fig:evaluation:sp:agent}
    \end{subfigure}
    \hfill
    \vspace{-0.3cm}
    \caption{[simulation] llama3-8B (Mooncake dataset)}
    \label{fig:evaluation:sp}
\end{figure}

\parab{End-to-End TTFT SLO Attainment.} We evaluate the end-to-end TTFT SLO attainment under varying request rates across diverse models and workloads. As shown in Fig~\ref{fig:evaluation:qwenA}, \ref{fig:evaluation:qwenB}, and \ref{fig:evaluation:sp}, \sys{} consistently outperforms state-of-the-art baselines across all scenarios.

Figure~\ref{fig:evaluation:qwenA} details the performance of \sys{} on conversation workloads for diverse MoE models. Specifically, under high request rates, \sys{} achieves $1.4\times$--$1.8\times$ higher SLO attainment for Mixtral (Fig.~\ref{fig:evaluation:qwenA:mixtral_200}), Qwen Coder (Fig.~\ref{fig:evaluation:qwenA:qwencoder_200}) and Grok (Fig.~\ref{fig:evaluation:qwenA:grok2_200}), and up to $2.4\times$ for DBRX (Fig.~\ref{fig:evaluation:qwenA:dbrx_200}). Furthermore, \sys{} can sustain $1.17\times$--$1.46\times$ higher request rates than the strongest baseline (Karuna) while maintaining the similar SLO attainment. In conversation workload, a small fraction of tail requests necessitates large KV-cache movements, which precipitate contention across Stage~1 to Stage~3 under high load. \sys{} outperform baselines significantly since it successfully coordinates multi-stage contention. Among the baselines, Karuna performs relatively better because it \emph{happens to} defer non-urgent Stage~3 traffic and provides partial relief; however, without explicit multi-stage coordination, it eventually succumbs to contention as other baselines.

Figure~\ref{fig:evaluation:qwenB} details the performance of \sys{} on agent workloads for diverse MoE models. Focusing on the mid-to-high load regime, \sys{} achieves $1.4\times$--$1.7\times$ higher SLO attainment for Grok-2 and Mixtral, and reaches $2.0\times$ for DBRX and Qwen-Coder. Furthermore, \sys{} sustains $1.2\times$ higher request rates for Grok-2, $\sim1.3\times$ for Mixtral and Qwen-Coder, and up to $1.4\times$ for DBRX compared to the Karuna. Compared to conversation traces, agent requests have shorter prompt length but exhibit higher average reuse rates, with multiple concurrent requests sharing identical prefixes. This pattern induces severe one-to-many sender-side contention, which primarily concentrates on Stage~1 and Stage~2 as multiple workers simultaneously contend for the same cached states. \sys{} effectively mitigates this contention by protecting urgent flows without requiring precise laxity. Notably, the performance gap between Karuna and other baselines narrows in this scenario. When contention is not dominated by Stage~3 traffic, Karuna falls back to SJF (deadline is implicit) to schedule Stage~1 and Stage~2 flows; however, being unaware of stage semantics, this approach still exposes tight requests to SLO violations as other baselines.

Figure~\ref{fig:evaluation:sp} details the performance of \sys{} on two representative long-context workloads using the Llama3-8B model with Sequence Parallelism (SP). For the Mooncake dataset, the conversation and agent workloads follow request patterns similar to those in the Qwen experiments (e.g., reuse ratios, access patterns), but feature significantly longer context lengths. In the conversation workload (Fig.~\ref{fig:evaluation:sp}a), \sys{} demonstrates robust performance under increasing load, achieving $1.3\times$--$1.6\times$ higher SLO attainment against Karuna. This advantage is further amplified in the agent workload(Fig.~\ref{fig:evaluation:sp}b), where \sys{} attains $1.4\times$--$1.9\times$ higher SLO attainment and supports up to $1.15\times$ higher per-gpu request rates given tight SLO budget.

\begin{figure}[t!]
    \centering
    \begin{subfigure}[t]{0.48\linewidth}
        \centering
        \includegraphics[width=\linewidth]{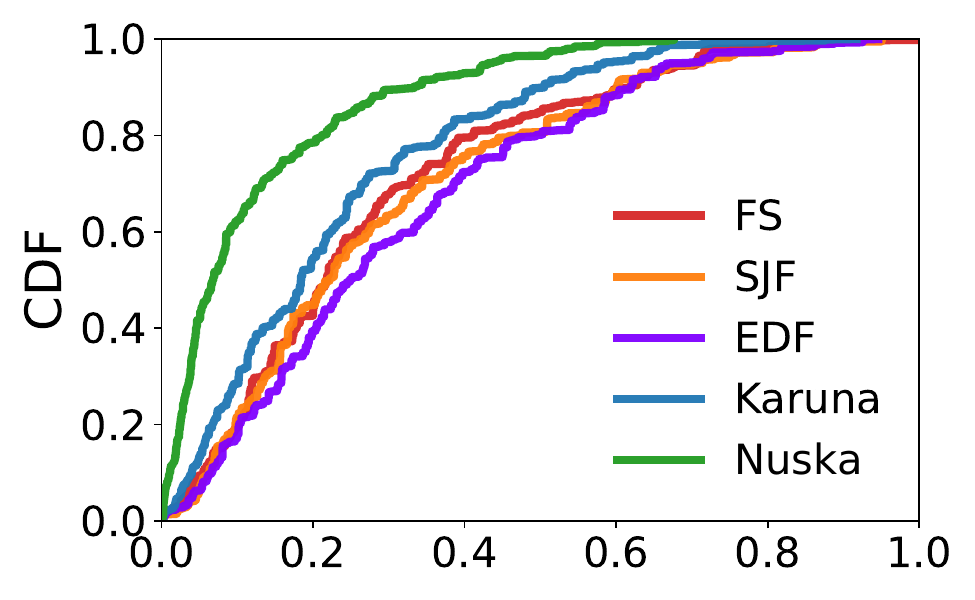}

        \caption{Normalized CCT.}
        \label{fig:evaluation:qwenA:dbrx:cascading}
    \end{subfigure}
    \hfill
    \begin{subfigure}[t]{0.48\linewidth}
        \centering
        \includegraphics[width=\linewidth]{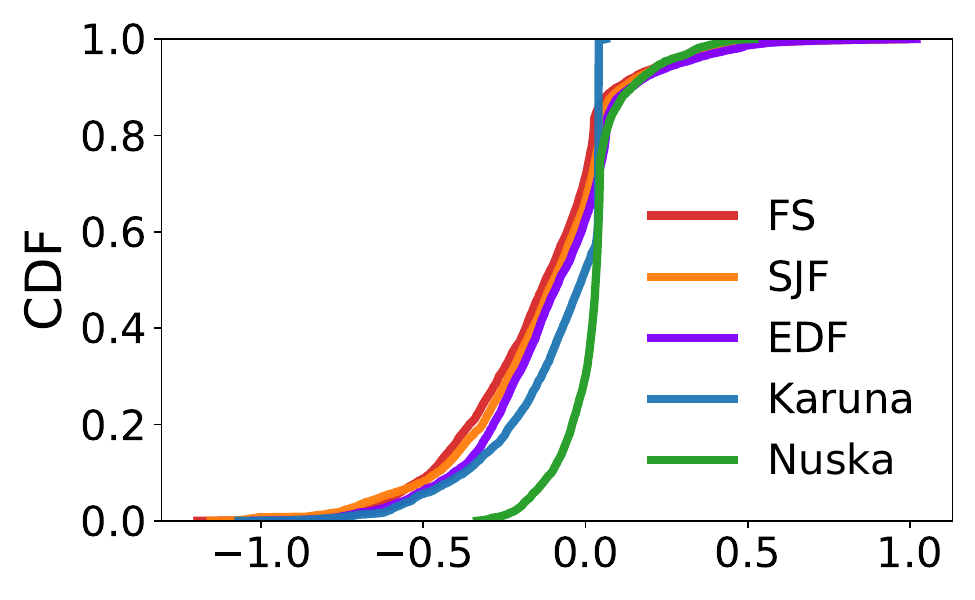}
        \caption{Normalized earliness.}
        \label{fig:evaluation:qwenA:dbrx:earliness}
    \end{subfigure}
    \hfill
    \caption{[Simulation] Breakdown on DBRX (Qwen-A, Rate=0.7)}
    \label{fig:evaluation:qwenA:dbrx:mb}

\end{figure}

\begin{figure}[t!]
    \centering
    \begin{subfigure}[t]{0.48\linewidth}
        \centering
        \includegraphics[width=\linewidth]{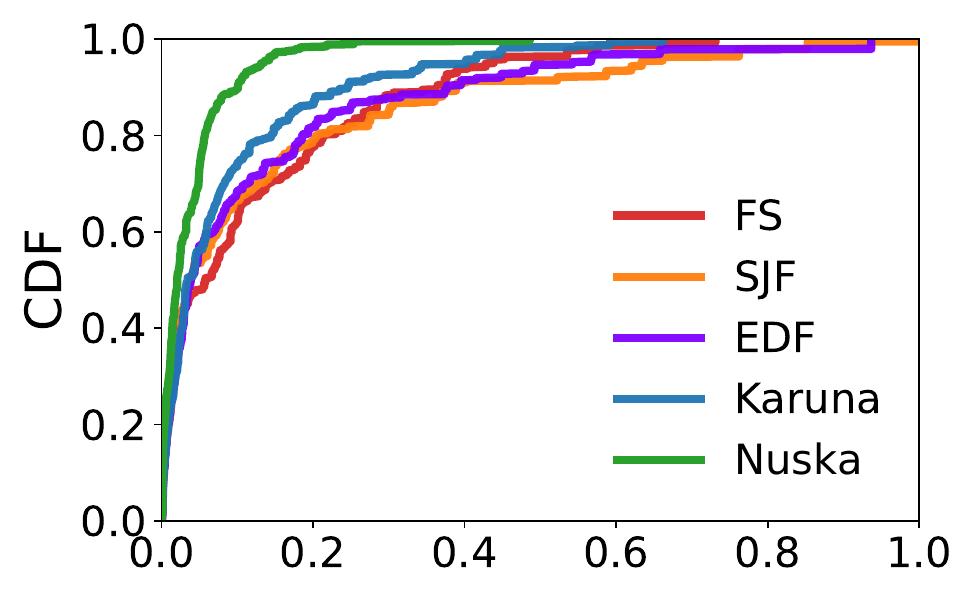}

        \caption{Normalized CCT.}
        \label{fig:evaluation:sp:cascading}
    \end{subfigure}
    \hfill
    \begin{subfigure}[t]{0.48\linewidth}
        \centering
        \includegraphics[width=\linewidth]{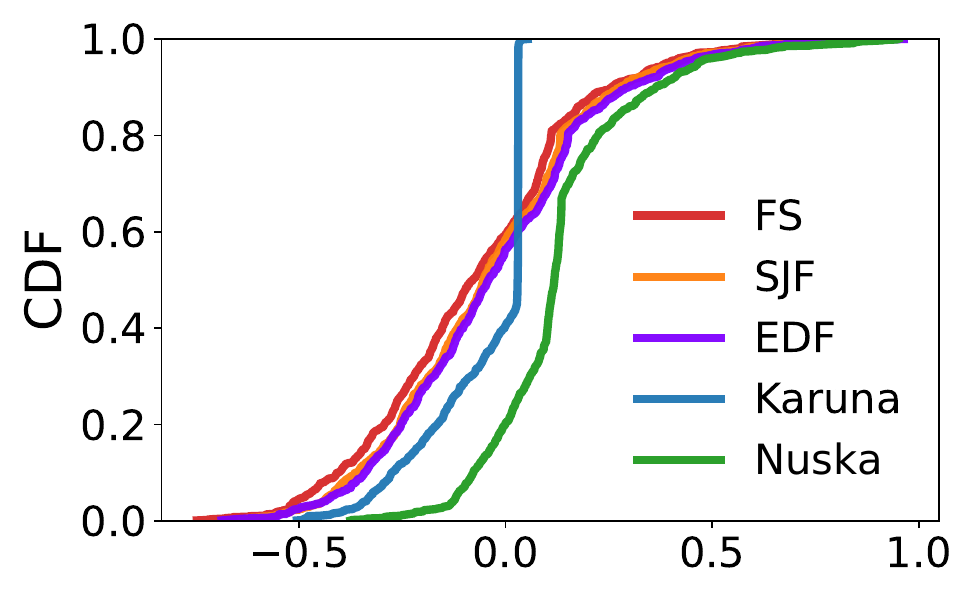}

        \caption{Normalized earliness}
        \label{fig:evaluation:sp:earliness}
    \end{subfigure}
    \hfill
    \caption{[Simulation] Breakdown on Llama3 (Mooncake-Conv, Rate=0.5)}
    \label{fig:evaluation:sp_mb}
    \vspace{-0.5em}
\end{figure}

\parab{Breaking down the performance gains.} We further conduct micro-benchmarks to analyze the source of the gains. As shown in Figure~\ref{fig:evaluation:qwenA:dbrx:mb} and Figure~\ref{fig:evaluation:sp_mb}, \sys{} successfully (i) minimizes Collective Completion Time (the non-overlapping communication latency) to shorten the prefill makespan, and (ii) regulates earliness to protect truly urgent requests under contention.

Figure~\ref{fig:evaluation:qwenA:dbrx:mb} breaks down the performance gains that lead to the $2.2\times$ higher SLO attainment achieved by \sys{} on the DBRX model.
The first source of improvement comes from reduced execution time during the prefill phase. Specifically, \sys{} lowers the average CCT of expert parallel by $52\%$ (Figure~\ref{fig:evaluation:qwenA:dbrx:cascading}). This reduction is mainly attributed to our stage-aware policy, which protects flows that block computation while deferring non-critical traffic to exploit available slack for overlap.
As a result, \sys{} significantly shortens the prefill makespan by lowering the effective communication overhead.

Beyond latency reduction, \sys{} further improves SLO attainment by better prioritizing truly urgent requests under contention.
We characterize this effect using \emph{earliness} of request, where large positive values indicate unnecessary early completion that can block urgent requests, while negative values correspond to deadline misses.
As shown in Figure~\ref{fig:evaluation:qwenA:dbrx:earliness}, Karuna exhibits the smallest non-negative earliness, but at the cost of a high violation risk, as it conservatively allocates near-minimal rates and fails to exploit available bandwidth.
In contrast, other baselines show much larger non-negative earliness, indicating that they schedule flows without effectively prioritizing urgency.
\sys{} reduces the non-negative portion of earliness by $42\%$ comparing with FS, SJF, EDF. By deferring Stage~3 flows to complete \textit{just-in-time} and regulating early-stage traffic with a RED-based ording, \sys{} preserves resources for turely urgent requests. \sys{} adopts a priority-based scheduling that can exploit available bandwidth when no higher-priority flows are pending.

We observe a similar trend in Figure~\ref{fig:evaluation:sp_mb}.
Across both micro-benchmarks, \sys{} consistently achieves higher SLO attainment by simultaneously reducing the collective completion time of sequence parallel (Figure~\ref{fig:evaluation:sp:cascading}) and regulating earliness under contention (Figure~\ref{fig:evaluation:sp:earliness}).
These results indicate that the performance gains of \sys{} are robust across models and are not tied to a specific workload or architecture.

\section{Discussion}

\parab{Additional parallelism.} While \sys primarily focuses on Expert Parallelism and Sequence Parallelism, we acknowledge other strategies not explicitly covered in our evaluation. Tensor Parallelism, widely adopted in serving, is typically confined to intra-node fabrics (e.g., NVLink) without directly contending for inter-node bandwidth. Similarly, we do not explicitly discuss Pipeline Parallelism, as the prefill stage is usually latency-sensitive (requiring low or zero PP degrees~\cite{distserve}), and its communication volume is orders of magnitude smaller than KV cache movement or collective communications~\cite{liao2025mixnet}. Should these flows appear on the shared network, they would be identified as Stage~2 traffic and remain compatible with our analysis. This observation extends to emerging parallelism strategies, which may introduce novel communication patterns with enhanced overlapping capabilities, yet still block dependent computation if delayed.

\parab{Hybrid model deployment.}
In practice, service providers often deploy a tiered model portfolio, consisting of a single large flagship model alongside multiple smaller models serving diverse workloads~\cite{openai2024gpt4}.
In such settings, network contention primarily arises from concurrent KV cache transfers (Stage~1 and Stage~3) across different models, as collective communication is typically confined within isolated racks or dedicated pods~\cite{deepseektech}.
While \sys{} is not explicitly evaluated under multi-model deployment, its scheduling principle remains applicable to KV cache traffic on shared networks.
While promising, the holistic orchestration of such multi-model contention landscapes remains an under-explored design space.

\parab{Applicability to scale-up fabrics.} While \sys focuses on scale-out networks, its design principles could be extended to scale-up domains (e.g., NVL~\cite{nvidia_gb200_nvl72}, UB~\cite{huawei384},etc.). Contention between collective and KV Cache movement remains when multiple serving instance deployed together. \sys's scheduling policy can be effectively adapted to such architectures by leveraging hardware-level isolation mechanisms (e.g., virtual channels or traffic classes), which could be exploited to enable differentiated prioritization.

\parab{Compatibility with GPU-initiated collective communication.} While emerging GPU-initiated communication~\cite{nvidia_nvshmem_ibgda,deepseek_deepep_2024} optimizes small-message latency for the decoding phase, it remains uncommon in prefill deployments due to strict hardware prerequisites and marginal gains for large-volume transfers. Nevertheless, even in such environments, \sys ensures compatibility via CPU-assisted variants (e.g., host-side doorbells), while this method may introduce negligible latency overhead compared to a fully GPU-initiated execution \cite{nvidia_magnum_nvshmem3}.

\section{Related Work.}

\parab{Flow and co-flow scheduling.}
Traditional flow scheduling optimizes per-flow metrics such as average flow completion time~\cite{dctcp,dcqcn,pfabric,PDQ,pias,li2017rate,li2024flow} or deadline satisfaction~\cite{D3,d2tcp,PDQ,zhang2015guaranteeing} using policies like SJF and EDF.
Coflow scheduling~\cite{varys,Aalo,zhang2016coda,zhao2015rapier,susanto2016stream,agarwal2018sincronia,wan2025coflow} extends this abstraction by grouping related flows and optimizing Coflow Completion Time for data-parallel workloads, while mixed-flow scheduling~\cite{karuna} considers the coexistence of deadline-sensitive and best-effort traffic. In contrast, \sys{} explicitly accounts for multi-stage dependencies to schedule flows toward end-to-end TTFT attainments.

\parab{Optimization for individual stages.} Prior research focuses on optimizing specific communication phases in isolation. For collective communication, efforts improve performance by algorithms synthesis~\cite{shah2023taccl,cao2025syccl}, resolving  potential contention~\cite{cao2024crux,rajasekaran2024cassini}, or enhancing overlap through system-level pipelining~\cite{comet,zheng2025triton,hong2025flashoverlap},. In parallel, other works optimize KV cache transfers by hiding latency via prefetching~\cite{mooncakefast,splitwise}, improving efficiency via block coalescing~\cite{li2025flowkv,chen2024kvdirect} and multi-path transmission~\cite{wu2026dualpath}, or reducing data volume by trading off model quality~\cite{cachegen}. However, these approaches treat each stage in isolation, overlooking the potential contention between collective communication and KV cache transfers.

\parab{LLM serving systems.} Recent advancements in LLM serving have evolved along two complementary dimensions. Architecturally, systems adopt prefill-decode disaggregation~\cite{distserve,splitwise} and KV-cache reuse~\cite{mooncakefast,cachedattention} to improve throughput and avoid redundant computation. On the orchestration front, numerous request schedulers~\cite{thunderserve,stojkovic2025dynamollm,chen2025slos,hongsola,tempo2025} have been proposed to maximize SLO attainment. However, these systems primarily focus on optimizing compute efficiency under ideal network assumptions. \sys is orthogonal to these approaches. It bridges the gap by addressing the network contention that they overlook.

\section{Conclusion}
This paper presented \sys{}, a holistic multi-stage flow scheduling framework for LLM serving that targets end-to-end TTFT SLO attainment. \sys{} is built on the observation that request-level deadlines progressively materialize into explicit flow level deadline as prefill execution. At its core, \sys{} realizes a Defer-and-Promote scheduling principle that approximates Least-Laxity-First behavior without requiring precise laxity estimation. Through a prototype implementation and extensive evaluation, we show that \sys{} consistently improves TTFT SLO attainment across diverse models and workloads.

\normalem

\bibliographystyle{ACM-Reference-Format}
\bibliography{reference}

\appendix

\section{Proof for Theorem 1}\label{sec:appendix:The1}

In this section, we provide the formal proof for the optimality of the \emph{Smallest-RLI-First} policy stated in Theorem 1.\subsection{Problem Setup and Definitions}We model the prefill process as a sequence of alternating communication and computation stages across layers $\ell = 1, \dots, N$.Let $L_{\text{curr}}(t)$ denote the layer index being computed or waiting to be computed at time $t$.For any communication flow $f$, let $L_{\text{target}}(f)$ be the layer where the data is consumed. The Relative Layer Index is defined as $\text{RLI}(f, t) = L_{\text{target}}(f) - L_{\text{curr}}(t)$.We adopt the assumptions stated in the theorem:\begin{enumerate}\item[(i)] \textbf{Ideal Computation:} Computation for layer $\ell$ starts immediately once its dependencies (flows with $L_{\text{target}} = \ell$) are met and runs for duration $C_\ell$ without interruption.\item[(ii)] \textbf{Fluid Model:} Communication flows are infinitely divisible, allowing preemption without overhead.\item[(iii)] \textbf{Dedicated Bandwidth:} We focus on a single link constraint (e.g., ingress or egress bottleneck).\end{enumerate}\subsection{Proof of Optimality}\textbf{Objective.} Minimizing the prefill makespan is equivalent to minimizing the completion time of the final computation layer. Since the total volume of computation is fixed, this is equivalent to minimizing the total duration of \emph{Execution Stalls} (intervals where the GPU is idle waiting for data).\textbf{Interval Classification.} We partition the timeline $[0, T]$ into two types of intervals based on the state of the system:\begin{itemize}\item \textbf{Stall Intervals ($\mathcal{I}_{\text{stall}}$):} Time periods where the GPU is idle because a flow $f$ with $\text{RLI}(f, t) = 0$ (i.e., targeting the current layer) is pending.\item \textbf{Overlap Intervals ($\mathcal{I}_{\text{overlap}}$):} Time periods where the GPU is actively performing computation. During these intervals, the link is available to transmit flows with $\text{RLI}(f, t) > 0$ (i.e., future layers).\end{itemize}\textbf{The Exchange Argument.}Suppose there exists an optimal schedule $\mathcal{S}^*$ that minimizes the makespan but violates the Smallest-RLI-First rule. This implies that at some time $t \in \mathcal{I}_{\text{stall}}$, the schedule assigns bandwidth to a flow $y$ with a larger RLI while a flow $x$ with a smaller RLI is pending.Specifically, since $t$ is a stall interval, there must be some pending flow $x$ with $\text{RLI}(x) = 0$ (blocking the current layer). The violation implies $\mathcal{S}^*$ serves flow $y$ where $\text{RLI}(y) > \text{RLI}(x) = 0$ during $[t, t+\delta]$.We construct a new schedule $\mathcal{S}'$ by applying an exchange:\begin{enumerate}\item \textbf{Swap:} In $\mathcal{S}'$, we assign the interval $[t, t+\delta]$ to flow $x$ instead of $y$.\item \textbf{Effect on Stall:} By serving $x$ earlier, the dependency for the current layer $L_{\text{curr}}$ is satisfied $\delta$ time units earlier (assuming $x$ was the bottleneck). This strictly reduces the duration of the current $\mathcal{I}_{\text{stall}}$ and advances the start of the next computation phase (and thus the next $\mathcal{I}_{\text{overlap}}$).\item \textbf{Feasibility of $y$:} Flow $y$ is displaced from $t$. However, since $\text{RLI}(y) > 0$, flow $y$ is not required until a future layer. The completion of $x$ triggers the start of computation $C_{L_{\text{curr}}}$, creating a new Overlap Interval. We can safely move the transmission of $y$ into this newly created Overlap Interval. Since $y$'s deadline is strictly later than $x$'s, this deferral does not violate any dependencies.\end{enumerate}\textbf{Conclusion.} The constructed schedule $\mathcal{S}'$ reduces the cumulative stall duration by transforming a portion of $\mathcal{I}_{\text{stall}}$ into $\mathcal{I}_{\text{overlap}}$. Repeating this exchange argument for all violations proves that the schedule strictly prioritizing the smallest RLI (serving $\text{RLI}=0$ flows immediately to minimize Stalls, and using Overlap intervals for $\text{RLI}>0$ flows) yields the minimum makespan. \qed

\section{Full algorithm for Robust inter-request scheduling} \label{sec:appendix:fullalg}

\begin{algorithm}[t!]
    \caption{Inter-request Scheduling}
    \small
    \label{alg:ETScheduling_Min}
    \begin{algorithmic}[1]
    \Procedure{InterScheduling}{$\mathcal{B}$, $M$, $B$}
        \FuncInput{$\mathcal{B}, M$: Batches and Traffic Matrices}
        \FuncInput{$B$: Global total drop budget}
        \FuncOutput{$\sigma, \mathcal{H}$}

        \State $S(\cdot) \gets 0$ \Comment{Interference from high-priority batches}
        \State $\mathcal{P} \gets \emptyset$; \quad $\mathcal{H} \gets \emptyset$

        \LineComment{Step 1: Sorting via Effective Deadline}
        \State $\sigma \gets \Call{SortByRED}{\mathcal{B}}$
        \State Precompute load vectors $L$ from $M$

        \For{$k \gets 1$ \textbf{to} $|\sigma|$}
            \State Let $i \gets \sigma[k]$
            \State $\mathcal{P} \gets \mathcal{P} \cup \mathcal{B}_i$ \Comment{Add to candidate pool}

            \State $\widehat{F}_i \gets \Call{EstFinishTime}{S, L_i}$
            \LineComment{Step 2: Feasibility Check}
            \While{$\widehat{F}_i > D^{Lo}_i$ \textbf{and} $|\mathcal{H}| < B$ \textbf{and} $\mathcal{P} \neq \emptyset$}
                \State $u^\star \gets \arg\max_u (S(u) + L_i(u))$ \Comment{Bottleneck}
                \LineComment{Step 3: selective pruning}
                \State $r^\star \gets \arg\max_{r \in \mathcal{P}} \text{Load of } r \text{ on } u^\star$

                \State $\mathcal{H} \gets \mathcal{H} \cup \{r^\star\}$; \quad $\mathcal{P} \gets \mathcal{P} \setminus \{r^\star\}$

                \If{$r^\star \in \mathcal{B}_i$} \Comment{Drop request in current batch}
                    \State $L_i(\cdot)\gets L_i(\cdot)-\ell_{r^\star}(\cdot)$
                \Else \Comment{Drop request in higher priority}
                    \State $S(\cdot)\gets S(\cdot)-\ell_{r^\star}(\cdot)$
                \EndIf

                \State $\widehat{F}_i \gets \Call{EstFinishTime}{S, L_i}$
            \EndWhile

            \State $S(\cdot) \gets S(\cdot) + L_i(\cdot)$ \Comment{Update}
        \EndFor

        \State \Return $\sigma, \mathcal{H}$
    \EndProcedure
    \end{algorithmic}
\end{algorithm}

The overall workflow of the scheduling algorithm is illustrated in Algorithm~\ref{alg:ETScheduling_Min}. In this section, we elaborate on the scheduling logic, providing the comprehensive mathematical formulation of priority assignment and further details on the practical implementation of latency estimation and enforcement.

This section details the robust scheduling algorithm designed to handle imprecise laxity and overload. The workflow operates in three phases triggered by batch arrival and departure events.

\parab{Step 1: Sorting via Robust Effective Deadline.}
To mitigate the "Piggyback Effect," the scheduler employs the Robust Effective Deadline ($RED$) metric defined in \S\ref{design:interreq}. The prioritization process executes in two stages. First, for each batch $\mathcal{B}$ with $n$ requests, the system sorts the requests internally by their deadlines $d_1 \le d_2 \le \dots \le d_n$. It then performs a linear scan to identify the partition point $k^* = \operatorname*{arg\,max}_{1 \le k < n} (d_{k+1} - d_k)$ that yields the maximal inter-request gap. This calculation dynamically separates the outliers (Tight Set) from the majority (Loose Set). Based on this partition, the scheduler computes the $RED$ score using the derived sub-batch deadlines and the outlier ratio $f$. Finally, all active batches are organized into a global priority queue sorted by ascending $RED$ values. This ensures that the dispatch order effectively filters out transient urgency spikes from isolated outliers while preserving the priority of genuinely urgent workloads.

\parab{Step 2: Worst-case Feasibility Check.} Following prioritization, we perform an admission control check to prevent the "Black Hole Effect." We estimate the completion time $\widehat{T}_i$ by strictly accumulating delays under a worst-case assumption (no overlap between batches). Computation latency is treated as deterministic, derived from the static computation graph of the Transformer model and offline profiling on the target hardware \cite{agrawal2024vidur,distserve,mooncake}. Communication latency is estimated by analyzing the batch's traffic matrix to identify the bottleneck port $p^*$ in the network fabric; the delay is calculated as the cumulative load on $p^*$ divided by its bandwidth. For dynamic architectures like Mixture-of-Experts (MoE) where token routing is runtime-dependent, we rely on historical routing statistics. Prior studies \cite{li2025optimizing} indicate that despite high noise in individual traffic matrices, the end-to-end latency prediction error typically remains within 20\%. This level of accuracy is sufficient for our scheduling granularity, allowing us to reliably enforce feasibility against the target deadline ($D^{Lo}_{\min}$).

\parab{Step 3: Selective Pruning and Soft Enforcement.} If a batch fails the feasibility check (i.e., $\widehat{T}_i > D^{Lo}_{\min}$), the system triggers a surgical pruning mechanism. We identify the specific requests within the batch that contribute the maximum load to the bottleneck port $p^*$ and iteratively remove them from the feasible set until the predicted latency meets the deadline. Crucially, this pruning employs a \emph{soft enforcement} strategy. Instead of immediately discarding the pruned requests, the scheduler demotes them to a "Scavenger" priority class within a dual-queue system. The Main Queue is serviced with strict priority, while the Scavenger Queue is processed opportunistically only when the network is detected to be idle or when actual runtime latencies are lower than the pessimistic estimates. This hybrid approach ensures system stability under worst-case predictions while maximizing throughput by reclaiming resources when congestion does not materialize.

\parab{Summary.} It is important to emphasize the design hierarchy within this workflow. The RED metric acts as the primary tie-breaker, governing the global dispatch order to maximize service-level objective (SLO) compliance under normal to moderate load. The feasibility check and selective pruning operate strictly as complementary safeguards, activated only during pathological overload to prevent resource exhaustion. This separation of concerns ensures that the scheduler remains stable and predictable—relying on robust prioritization for the vast majority of decisions—while retaining the capability to degrade gracefully only when extreme contention makes it unavoidable.

\end{document}